\documentclass[
  twocolumn,
  pre,
  preprintnumbers,
  floatfix,
  superscriptaddress,
  citeautoscript, longbibliography,
]{revtex4-1}

\usepackage{physics}
\usepackage{amsmath}
\usepackage{amssymb}
\usepackage[dvipsnames]{xcolor}
\usepackage{graphicx}
\usepackage{tabularx}
\usepackage{subfigure}
\usepackage{dcolumn}
\usepackage{times}
\usepackage{comment} 
\usepackage{ulem}
\usepackage{url}

\newcommand{\sect}[1]{Sec.~\ref{#1}}
\newcommand{\eq}[1]{Eq.~(\ref{#1})}

\renewcommand{\braket}[3]{\bigl{<}#1\big|#2\big|#3\bigr{>}}
\renewcommand{\bra}[1]{\bigl{<}#1\big|}
\renewcommand{\ket}[1]{\big|#1\bigr{>}}
\newcommand{\ibraket}[2]{\bigl{<}#1\big|#2\bigr{>}}

\begin{document}

\title{
Spectral-partitioned Kohn-Sham density functional theory
}

\author{Babak Sadigh}
\email{sadigh1@llnl.gov}
\author{Daniel \AA berg}
\author{John Pask}
\affiliation{Lawrence Livermore National Laboratory, Livermore, CA, 94550}

\date{\today}

\begin{abstract}
  We introduce a general, variational scheme for systematic approximation of a given Kohn-Sham free-energy functional by partitioning the density matrix into distinct spectral domains, each of which may be spanned by an independent diagonal representation without requirement of mutual orthogonality. It is shown that by generalizing the entropic contribution to the free energy to allow for independent representations in each spectral domain, the free energy becomes an upper bound to the exact (unpartitioned) Kohn Sham free energy, attaining this limit as the representations approach Kohn-Sham eigenfunctions. A numerical procedure is devised for calculation of the generalized entropy associated with spectral partitioning of the density matrix. The result is a powerful framework for Kohn-Sham calculations of systems whose occupied subspaces span multiple energy regimes. As a case in point, we apply the proposed framework to warm- and hot-dense matter described by finite-temperature density functional theory, where at high energies the density matrix is represented by that of the free-electron gas, while at low energies it is variationally optimized. We derive expressions for the spectral-partitioned Kohn-Sham Hamiltonian, atomic forces, and macroscopic stresses within the projector-augmented wave (PAW) and the norm-conserving pseudopotential methods. It is demonstrated that at high temperatures, spectral partitioning facilitates accurate calculations at dramatically reduced computational cost. Moreover, as temperature is increased, fewer exact Kohn-Sham states are required for a given accuracy, leading to further reductions in computational cost. Finally, it is shown that standard multi-projector expansions of electronic orbitals within atomic spheres in the PAW method  lack sufficient completeness at high temperatures. Spectral partitioning provides a systematic solution for this fundamental problem.
\end{abstract}

\maketitle

\section{Introduction}

Most complex problems in materials chemistry and physics have heterogeneous character involving many length, time, and energy scales. Often solutions exist for separate spatial, temporal, or spectral domains but difficulties arise when they are merged while boundary interactions are accounted for and global constraints are maintained. In electronic structure theory, there are many instances of this approach, e.g., the pseudopotential approximation for the core-valence interaction \cite{BHS,Ihm_1979,Teter_RMP,ultrasoft}, the divide-and-conquer technique for order-{\it N} scaling density functional theory (DFT) \cite{Yang_PRL_91,LinWang_D&C}, the coherent potential approximation for disordered alloys \cite{CPA}, and the downfolding technique in many-body physics for strongly correlated electrons embedded in a Fermi liquid \cite{FerdiPRL2009,RMP_kotliar}.

The aim of this paper is to introduce and develop a general variational framework for spectral partitioning (SP) of the density matrix (DM) in Kohn-Sham DFT. In this scheme, a spectral-partitioned DM is constructed from independent diagonal representations, each spanning a  distinct energy domain. This is accomplished by using a spectral partition of unity to ensure that the original unpartitioned DM is recovered in the limiting case that the representations in all subdomains consist of Kohn-Sham (KS) eigenfunctions. 

An important example of spectral partitioning in chemistry is the subdivision of the occupied subspace into core and valence states. The core states are treated as localized atomic-like orbitals, while the valence electrons are allowed to become extended with Bloch wave character and required to be orthogonal to the core subspace \cite{LAPW_Andersen,FPLAPW_Wimmer}. As a result of this orthogonality constraint, the valence wavefunctions in molecules and solids become quite complex, exhibiting rapid spatial variations near the nuclei and bond formation in the interstitial region between the nuclei. By relaxing the orthogonality constraint, the pseudopotential approximation in its various forms \cite{JCPhillips, BHS, ultrasoft, ONCV, Bloechl}  achieves a much simpler description of the valence subspace. The separate treatments of the two subspaces is possible due to the substantial energy gap between them. Our purpose in this paper is to develop a robust and general framework, which we refer to as spectral-partitioned Kohn-Sham density functional theory (spDFT), enabling the use of different representations in different energy ranges, for any electronic structure, regardless of presence or size of energy gaps. 

While spectral partitioning is a general mathematical technique applicable to the full range of electronic structure problems, as well as generalizations of KS theory, we have been motivated by problems that plague finite-temperature DFT calculations of high-energy-density (HED) matter. These calculations play a significant role in the fundamental understanding of exciting new fields of physics, from inertial confinement fusion \cite{ICF} to laboratory astrophysics \cite{astro}, that have emerged due to recent advances in laser and pulsed power technologies \cite{NIF, LCLS, Omega}. The conditions achieved in HED experiments are so complex and so difficult to characterize that theory and computations are indispensable for both their design and interpretation. Calculations of equations-of-state (EOS), opacities, and X-ray absorption spectra are but a few examples of necessary contributions from theory.  Finally, as a result of these developments, significant progress has been achieved in our understanding of the structures of planetary interiors and their magnetic fields \cite{planets}.

Several complications arise when standard implementations of KS-DFT are applied to HED matter. At extreme temperatures, a substantial density of highly excited nearly-free electrons coexist with low-energy hybridizing valence electrons as well as ionized core shells \cite{surhbarbee}.  Hence, a large number of highly excited states must be incorporated in the calculations, which can lead to prohibitive computational costs. In order to circumvent this so-called ``orbitals wall'' problem, a variety of approaches and approximations have been employed, including orbital-free approaches \cite{OFDFT_Nagy, OFDFT_Trickey1, OFDFT_Karasiev_2020}, density-matrix based techniques \cite{SQ1, SQ2, SQ3}, Green's function methods \cite{Starett}, path-integral Monte Carlo \cite{PIMD1, PIMD2}, and pseudo-atom molecular-dynamics \cite{PAMD1, PAMD2}. Each of these techniques has its advantages and disadvantages. In this regard, spectral partitioning provides an interesting alternative as it can alleviate the orbitals-wall problem by breaking up the valence-electron subspace into two spectral domains: (i) the low-energy subspace of hybridizing orbitals, which can be treated by exact diagonalization of the KS Hamiltonian, and (ii) the high-energy subspace of highly excited states, which can be treated as nearly-free electron states. In this way, computational cost can be reduced dramatically without loss of accuracy. Among other HED conditions of interest are extreme densities. At these conditions, core levels can overlap and form bands, and the energy window within which hybridization occurs can become exceedingly wide. Spectral partitioning can offer an effective solution to this problem by splitting the spectral range of the occupied subspace into smaller intervals, each of which is treated separately and merged seamlessly.

Zhang and coworkers  \cite{Zhang} have recently pioneered the idea of spectral partitioning at extreme temperatures by splitting the expectation values of relevant observables into exact KS contributions at low energies and approximate homogeneous electron gas (HEG) contributions at high energies, the so-called extended first-principles molecular-dynamics (ext-FPMD) method. They implemented a self-consistent scheme within the PAW method, and have shown much promise for calculations of plasma EOS. This methodology has been further developed and employed in subsequent works \cite{Abinit, Abinit1, Abinit2, Hollebon, LiuZhang}.

However, all ext-FPMD formulations to date rely on intuition and approximation regarding key aspects such as the coupling of KS and HEG contributions and handling of nonlocal pseudopotentials. This in turn is due to the lack of a variational free-energy functional from which the ext-FPMD Hamiltonian, forces, and stresses can be analytically derived. Specific consequences include: (i) {\it ad hoc} expression for the self-consistent Hamiltonian, (ii) inconsistency between the expression for free energy and those for forces and stresses, and (iii) internal inconsistency between forces/stresses calculated at different points along ionic trajectories.

In this paper, we show how all of the issues listed above can be straightforwardly and rigorously addressed using the spDFT framework. We show that a variational spDFT free-energy functional can be derived for any non-pathological spectral decomposition of the DM. The key innovation is to generalize the entropy function to allow for independent representations, other than just KS eigenfunctions, in each spectral domain. The need for amending total-energy functionals with entropy terms was first realized when generalizing KS-DFT to finite temperatures \cite{Mermin,Karasiev_PRL1,Karasiev_PRL2,GDB_PRL}, and was later found to be essential for internal consistency of ab-initio total energies and atomic forces when smearing techniques are used to carry out Brillouin-zone (BZ) integrations \cite{Ho_Gaussian,Meth-Paxt,Pedersen91,Wentz92,Davenport92,Marzari_PRL97,Marzari_PRB23}. The derivation in this work of the entropy associated with spectral partitioning of the DM builds on this foundation, and introduces further technical advances to it, as we detail below. 

With the SP-entropy in place, we show that the total SP free energy is an upper bound to the exact (unpartitioned) KS free energy. Consequently, self-consistent Hamiltonians, as well as expressions for forces and stresses, can be straightforwardly derived from the variational principle. Furthermore, the variational spDFT free-energy functional can now be endowed with higher order corrections via perturbation theory \cite{Baroni, deGironcoli, Gonze89}, or generalized to other contexts that may benefit from spectral partitioning with a strongly inhomogeneous electron gas, for which intuitive guesses become inadequate and a rigorous variational framework as developed here becomes indispensible. 

In the following, we first formulate a general framework for spectral partitioning of the DM using an analytic partition of unity to smoothly combine different representations in distinct energy domains. Subsequently, an associated spDFT free-energy functional is constructed, from which forces, stresses, and related physical quantities can be derived. We then discuss practical algorithms for convenient and user-friendly implementations of the spDFT framework that can handle elaborate Fermi surfaces, and are able to maintain consistency between total energy and forces throughout dynamical simulations. To illustrate the power of the framework in practice, we develop the detailed formalism for incorporation of the HEG approximation at high energies, and derive expressions for its implementation within PAW and norm-conserving pseudopotential (NCPP) techniques. We then apply the new methodology to the study of H and Be lattices at warm dense matter and plasma conditions. As an unexpected outcome of this study, we show that at elevated electron temperatures, standard nonlocal projector expansions for pseudo-wavefunctions become increasingly incomplete within the atomic augmentation spheres, and as a result, the basic assumptions underlying the derivation of the PAW equations break down. We discuss the ubiquity of this problem and demonstrate how spDFT can be used to rectify it.

\section{\lowercase{sp}DFT Derivation}
\label{sec:derivation}

Spectral partitioning is a general technique that can be applied not only to finite-temperature DFT and local KS functionals \cite{PBE,Karasiev_PRL1,Karasiev_PRL2}, but also to generalized KS functionals  such as meta-GGA \cite{metaGGA_PRL15}, DFT+U \cite{Anisimov97,Coco05} and hybrid functionals \cite{B3LYP,PBE0, HSE,Truhlar}. Furthermore, spectral partitioning can be applied to any electronic occupation statistics, such as Gaussian smearing \cite{Ho_Gaussian} or Fermi-Dirac broadening \cite{Mermin}. However, for clarity and brevity, we focus in this paper on application of spectral partitioning to finite-temperature KS-DFT for extended systems in periodic boundary conditions. In the following, Hartree atomic units are used unless otherwise specified.

\subsection{Ensemble Kohn-Sham density functional theory}

Consider a many-electron system in an external ionic potential  $\hat{V}_{ie}(\{{\bf R}\})$, with the nuclei at positions $\{{\bf R}\}$ in a periodic array of unit cells each containing $N_{at}$ atoms in a volume $\Omega$. Ensemble Kohn-Sham DFT maps this system onto a reference system of non-interacting electrons in an external self-consistent potential $\hat{V}_{KS}$. The state of the non-interacting system is completely described by the ensemble density operator $\hat{\rho}$, whose real-space representation is the density matrix $\rho({\bf r'},{\bf r})$ defined as
\begin{eqnarray}
  \label{eq:rho}
  \rho({\bf r'},{\bf r})&\equiv&\braket{{\bf r'}}{\hat{\rho}}{{\bf r}} = \sum_{{\bf k},n} f_{{\bf k}n} \psi_{{\bf k}n}({\bf  r}) \psi^*_{{\bf k}n}({\bf r'}), 
\end{eqnarray}
where the ${\bf k}$-index enumerates $N_{BZ}$ Bloch wave vectors on a uniform grid spanning the first BZ, $n$ enumerates bands, $\psi_{{\bf k}n}({\bf r})$ are the KS wavefunctions, and $f_{{\bf k}n}$ are occupation probabilities required to be non-negative, $f_{{\bf k}n} \ge 0$, and are derived from ensemble statistics, as shown below. The discrete grid of Bloch wave vectors derives from the Born von Karman (BvK) boundary condition on the wavefunctions. Within the BvK supercell, every pair of Bloch wavefunctions with wave vectors ${\bf k} \ne {\bf k'}$ are orthogonal. Furthermore, finite systems can be represented with periodic boundary conditions in the limit of vanishing density of ions in each unit cell.

In the following, the domain of all integrals is the unit cell volume $\Omega$ unless otherwise specified. Also, we define the Bloch wavefunctions as
\begin{equation}
  \label{eq:psi-u}
 \psi_{{\bf k}n}({\bf r}) = \frac{1}{\sqrt{N_{BZ}}} u_{{\bf k}n}({\bf r}) e^{i{\bf k}\cdot{\bf r}},
\end{equation}
where $u_{{\bf k}n}$ have the periodicity of the lattice and are normalized in the unit cell. 

As a result of Bloch's theorem, the density matrix $\hat{\rho}$ can be decomposed into Bloch-wave components
\begin{eqnarray}
  \label{eq:rhok}
\braket{{\bf r'}}{\hat{\rho}}{{\bf r}} = \frac{1}{N_{BZ}}\sum_{\bf k} \braket{{\bf r'}}{\hat{\rho}_{\bf k}}{{\bf r}} e^{i{\bf k}\cdot({{\bf r}-{\bf r'}})}, 
\end{eqnarray}
with the property that every pair of Bloch-wave components $\hat{\rho}_{\bf k}$ and $\hat{\rho}_{\bf k'}$ with ${\bf k}\ne{\bf k'}$ are orthogonal in the BvK supercell, i.e.,
\begin{equation}
\label{eq:orthodm}
\int_{\text{BvK}} \braket{{\bf r'}}{\hat{\rho}_{\bf k}}{{\bf r}} \braket{{\bf r}}{\hat{\rho}_{\bf k'}}{{\bf r''}}e^{i({\bf k}-{\bf k'})\cdot{\bf r}}~d{\bf r} = 0
\end{equation}  
for all ${\bf r'}$ and ${\bf r''}$ in the BvK cell. Hence, the KS problem is separable with respect to the lattice-periodic operators ${\hat{\rho}_{\bf k}}$, 
\begin{equation}
 \label{eq:rhokkk}
\braket{{\bf r'}}{\hat{\rho}_{\bf k}}{{\bf r}} =  \sum_{n} f_{{\bf k}n} u_{{\bf k}n}({\bf  r}) u^*_{{\bf k}n}({\bf r'}), 
\end{equation}
which contain all the variational degrees of freedom.


The charge density $n({\bf r})$ is obtained as the diagonal of the DM
\begin{equation}
  \label{eq:n}
  n({\bf r}) = \rho({\bf r},{\bf r}).
\end{equation}

The Helmholtz free energy of the non-interacting system in the absence of the self-consistent potential $\hat{V}_{KS}$ can be written
\begin{eqnarray}
\label{eq:6}
  \mathcal{A}_{KS}[\hat{\rho};\tau_e] = T_s[\hat{\rho}] - \tau_e \Tr\left\{S[\hat{\rho}]\right\},
\end{eqnarray}
where the non-interacting kinetic energy $T_s[\hat{\rho}]$ takes on the form
\begin{eqnarray}
  T_s[\hat{\rho}] &=& 
 -\frac{1}{2N_{BZ}}\sum_{{\bf k},n}f_{{\bf k}n} \braket{u_{{\bf k}n}} {({\bf \nabla}+i{\bf k})^2}{u_{{\bf k}n}},
\end{eqnarray}
and the entropy function $S[\hat{\rho}]$ is specified by the ensemble statistics and $\tau_e$ is the associated temperature. Note that the $\Tr$ operator in \eq{eq:6} corresponds to integration over a single unit cell and therefore we have
\begin{equation}
  \Tr\left\{S[\hat{\rho}]\right\} = \frac{1}{N_{BZ}}\sum_{{\bf k},n} S[f_{{\bf k}n}].
\end{equation}

With the non-interacting free energy $\mathcal{A}_{KS}$ defined, the total ensemble-KS free-energy $F_{KS}$ can be written as a functional of the DM $\hat{\rho}$ at temperature $\tau_e$ in an external potential  $\hat{V}_{ie}(\{{\bf R}\})$ 
\begin{align}
    \label{eq:En0}
    F_{KS}[\hat{\rho};\{{\bf R}\},\tau_e] & = E_{KS}[\hat{\rho};\{{\bf R}\},\tau_e]
    - \tau_e \Tr\left\{S[\hat{\rho}]\right\} \\
    & = \mathcal{A}_{KS}[\hat{\rho};\tau_e] +E_H[n] + F_{xc}[\hat{\rho},\tau_e] + \nonumber\\
  & \int \braket{{\bf r}}{\hat{V}_{ie}(\{{\bf R}\})}{{\bf r'}} \rho({\bf r'},{\bf r})~d{\bf r}d{\bf r'} \nonumber
\end{align}
where
\begin{eqnarray}
\label{eq:hart000}
E_H[n] &=& \frac{1}{2}\sum_{\bf T}\int \frac{n({\bf r})n({\bf r'})}{|{\bf r}-{\bf r'}+{\bf T}|} ~d{\bf r}d{\bf r'},\\
\label{eq:Vie}
\braket{{\bf r}}{\hat{V}_{ie}(\{{\bf R}\})}{{\bf r'}} &=&
\sum_{{\bf R}} V_{ie}({\bf r}-{\bf R},{\bf r'}-{\bf R}),\\
\label{eq:V_ie}
V_{ie}({\bf r},{\bf r'}) &=& V_{loc}({\bf r}) \delta({\bf r}-{\bf r'}) + V_{NL}({\bf r},{\bf r'}).
\end{eqnarray}
Above, $\hat{V}_{ie}(\{{\bf R}\})$ is the electron-ion interaction potential operator that is allowed to be nonlocal within each atomic sphere in case ions are replaced by pseudopotentials. In \eq{eq:hart000}, ${\bf T}$ denotes the set of periodic lattice translation vectors, and in \eq{eq:Vie}, the nuclear positions are expanded as ${\bf R} = {\bf T} + {\bf s}_i$, where ${\bf s_i}$ specify the positions of atoms within each unit cell. In \eq{eq:En0}, $F_{xc}[\hat{\rho},\tau_e]$ is the contribution of the electronic exchange and correlation (XC) to the free energy, which in general has explicit temperature dependence \cite{Karasiev_PRL1, Karasiev_PRL2, GDB_PRL}. Within the most commonly used approximations in KS-DFT \cite{PBE}, $F_{xc}$ is a functional of the diagonal elements of the DM only, i.e., charge density $n({\bf r})$ and its gradients, but XC functionals with explicit dependences on off-diagonal elements of the DM, such as hybrid exchange \cite{PBE0, HSE, Karasiev_hybrid} and meta-GGA \cite{metaGGA_PRL15, Karasiev_metaGGA} are becoming increasingly popular.

The formalism developed in the following is general. However, for the sake of illustration, we will focus on application to warm-dense matter, where a large number of partially occupied orbitals must be accounted for whose occupation probabilities $f_{{\bf k}n}$ are distributed according to Fermi-Dirac (FD) statistics. Below, we derive the associated entropy function $S[\hat{\rho}] = S^{FD}[\hat{\rho}]$. Generalization to other statistical ensembles is straightforward.

Let us start by formulating the expression for the equilibrium ensemble-KS free-energy $\Omega_{KS}$. This can be obtained by constrained minimization of $F_{KS}$ with respect to $\hat{\rho}$, which involves the variational degrees of freedom $\{u_{{\bf k}n}\}$ and $\{f_{{\bf k}n}\}$
\begin{eqnarray}
 \label{eq:freeen_exact}
  \Omega_{KS}[\{{\bf R}\},\tau_e] &=& \min_{\hat{\rho},\mu,\{\Lambda\}} ~~~ F_\text{KS}[\hat{\rho};\{{\bf R}\},\tau_e] \\ 
  &-&   \mu \left(\Tr\{\hat{\rho}\} - N_e\right)  \nonumber \\
  &-& \sum_{{\bf k},n,m} \Lambda_{nm}^{\bf k} \left(\left<u_{{\bf k}n}|u_{{\bf k}m}\right> -\delta_{nm}\right). \nonumber
\end{eqnarray}
Above, the second term on the right-hand side constrains the total number of electrons $N_e$, the last term enforces orthonormalization of the KS wavefunctions, and $\mu$ and $\Lambda_{nm}^{\bf k}$ are the associated Lagrange multipliers. It should be noted that within this formulation, no constraints are imposed on the XC potential to be multiplative and local. Hence, for XC functionals that depend explicitly on off-diagonal elements of the DM, such as meta-GGA and hybrid-exchange, the above procedure leads to semi-local or nonlocal XC potentials, which result in self-consistent Hamiltonians that belong to the generalized-KS framework \cite{GKSPNAS, GarrickPRX}. Inclusion of potential constraints that enforce rigorous KS mapping to fictitious non-interacting systems in a self-consistent optimized effective potential (OEP) \cite{Gorling05} are not considered in the present work. While spectral partitioning can in principle provide a powerful way to simplify the OEP integro-differential equations, the tools developed in this work are not immediately applicable to that problem.

At equilibrium, the KS wavefunctions $\psi_{{\bf k}n}$ become eigenfunctions of the self-consistent Hamiltonian $\hat{H}^{KS}$, which can be obtained through functional differentiation of $E_{KS}$ in \eq{eq:En0} with respect to $\hat{\rho}$ defined in \eq{eq:rho}:
\begin{equation}
\label{eq:dEdrho}
\braket{{\bf r'}}{\hat{H}^{KS}[\hat{\rho}^{KS}]}{{\bf r}} = \frac{\delta E_{KS}[\hat{\rho}^{KS}]}{\delta \rho({\bf r'},{\bf r})},
\end{equation}
where for brevity, we have suppressed the dependence of $E_{KS}$ and $\hat{H}^{KS}$ on $\{{\bf R}\}$ and $\tau_e$. The differentiation in \eq{eq:dEdrho}, in the BvK supercell, leads to 
\begin{eqnarray}
  \label{eq:Hamil}
\frac{\delta E_{KS}}{\delta \rho({\bf r'},{\bf r})} &=& \delta({\bf r}-{\bf r'})\left(-\frac{{\bf \nabla}^2}{2} + V_H({\bf r})\right) + \\
&& \braket{{\bf r'}}{\hat{V}_{ie}[{\bf R}]}{{\bf r}} + 
\braket{{\bf r'}}{\hat{V}^{xc}[\hat{\rho}^{KS},\tau_e]}{{\bf r}},\nonumber
\end{eqnarray}
with $V_H({\bf r}) = \delta E_H/\delta n({\bf r})$, $\braket{{\bf r'}}{\hat{V}_{ie}[\{{\bf R}\}}{{\bf r}}$ defined in \eq{eq:Vie}, and 
\begin{equation}
\label{eq:Vxc}
\braket{{\bf r'}}{\hat{V}^{xc}[\hat{\rho}^{KS},\tau_e]}{{\bf r}} = \frac{\delta F_{xc}[\hat{\rho}^{KS},\tau_e]}{\delta \rho({\bf r'},{\bf r})},
\end{equation}
where we have explicitly presented the dependences of the various potentials on the temperature $\tau_e$, ionic positions $\{{\bf R}\}$, and the DM $\rho^{KS}$. Note that for XC functionals that only depend on the density $n({\bf r})$ and its gradients, the XC potential becomes multiplicative and local: $\braket{{\bf r'}}{\hat{V}^{xc}[n,\tau_e]}{{\bf r}} = \delta({\bf r}-{\bf r'})\delta F_{xc}[n,\tau_e]/\delta n({\bf r})$.

At equilibrium, the Lagrange multiplier matrices $\Lambda_{nm}^{\bf k}$ become diagonal,
\begin{equation}
\label{eq:Lambda_nm}
  \Lambda_{nm}^{\bf k} = \delta_{nm}~ \frac{f_{{\bf k}n}}{N_{BZ}} \braket{u_{{\bf k}n}}{\hat{H}^{KS}_{\bf k}}{u_{{\bf kn}}} = \delta_{nm}~\frac{f_{{\bf k}n}}{N_{BZ}} \epsilon_{{\bf k}n},
\end{equation}
where $\hat{H}^{KS}_{\bf k}$ are lattice-periodic Bloch-wave components of the KS Hamiltonian defined as 
\begin{eqnarray}
  \label{eq:H-Hk}
  \braket{{\bf r'}}{\hat{H}^{KS}}{{\bf r}} &=& \frac{1}{N_{BZ}}\sum_{\bf k}\braket{{\bf r'}}{\hat{H}^{KS}_{\bf k}}{{\bf r}} e^{i{\bf k}\cdot({\bf r}-{\bf r'})},
\end{eqnarray}
and can be obtained through functional differentiation of $E_{KS}$ in \eq{eq:En0} with respect to $\hat{\rho}_{\bf k}$ defined in \eq{eq:rhokkk}:
\begin{equation}
\label{eq:dEdrho-k}
\braket{{\bf r'}}{\hat{H}^{KS}_{\bf k}[\hat{\rho}^{KS}]}{{\bf r}} = N_{BZ}\frac{\delta E_{KS}[\hat{\rho}^{KS}]}{\delta \rho_{\bf k}({\bf r'},{\bf r})}.
\end{equation}

Finally, at equilibrium the occupation probabilities become solutions to the equations
\begin{equation}
  \label{eq:fkn}
  \epsilon_{{\bf k}n} - \mu - \tau_e\frac{\partial S}{\partial f_{{\bf k}n}} = 0,
\end{equation}
where $\mu$ acts as the chemical potential. The left-hand side of the above equation is obtained by partial differentiation of the free energy expression \eq{eq:En0} with respect to occupations $f_{{\bf k}n}$. Within FD statistics, the occupation probabilities are distributed according to
\begin{equation}
  \label{eq:fFD}
  f^{FD}_{{\bf k}n} = \frac{1}{1 + \exp\left(\frac{\epsilon_{{\bf k}n} - \mu}{\tau_e}\right)}.
\end{equation}
By inserting \eq{eq:fkn} into \eq{eq:fFD}, a relation for the FD entropy function $S^{FD}[\hat{\rho}]$ can be obtained
\begin{equation}
  \label{eq:Sdot_fd}
  f_{{\bf k}n} = \frac{1}{1 + \exp\left(\dot{S}^{FD}(f_{{\bf k}n})\right)}, 
\end{equation}
with $\dot{S}^{FD}(f) = \frac{dS^{FD}}{df}$. Equation~(\ref{eq:Sdot_fd}) can be solved analytically for $\dot{S}^{FD}$, and integrated to obtain the FD entropy function $S^{FD}$ using the boundary condition $S^{FD}(0) = 0$, which subsequently can be written as
\begin{equation}
  S^{FD}\left(f_{{\bf k}n}\right) = -f_{{\bf k}n}~\ln(f_{{\bf k}n}) - (1-f_{{\bf k}n})~\ln(1-f_{{\bf k}n})).
  \label{eq:FDentropy}
\end{equation}

By inserting the FD occupations \eq{eq:fFD} into \eq{eq:rho}, and using the diagonal representation of the KS Hamiltonian 
\begin{eqnarray}
  \label{eq:H^KS}
  \hat{H}^{KS} &=& \sum_{{\bf k},n} \epsilon_{{\bf k}n} \ket{\psi_{{\bf k}n}}\bra{\psi_{{\bf k}n}},
\end{eqnarray}
the following relation can be derived between the equilibrium DM and the KS Hamiltonian operators
\begin{equation}
  \label{eq:exact}
  \hat{\rho}^{KS} = \left(\hat{I} + \exp\left(\frac{\hat{H}^{KS} - \mu \hat{I}}{\tau_e}\right)\right)^{-1},
\end{equation}
where $\hat{I}$ is the identity operator. Likewise, using the diagonal representation of $\hat{H}^{KS}_{\bf k}$
\begin{eqnarray}
  \hat{H}_{\bf k}^{KS} &=& \sum_{n} \epsilon_{{\bf k}n} \ket{u_{{\bf k}n}}\bra{u_{{\bf k}n}},
\end{eqnarray}
we can derive the following operator relation
\begin{equation}
  \label{eq:exact}
  \hat{\rho}^{KS}_{\bf k} = \left(\hat{I} + \exp\left(\frac{\hat{H}^{KS}_{\bf k} - \mu \hat{I}}{\tau_e}\right)\right)^{-1}.
\end{equation}
   

\subsection{spDFT at extreme temperatures}

At high electron temperatures the FD distribution becomes broad with a long spectral tail leading to finite occupations at very high energies $\epsilon_{{\bf k}n}$, which in turn makes orbital-based electronic-structure calculations in the regime of warm-dense matter computationally very expensive. One way around this problem has been suggested by Zhang {\it et al.} \cite{Zhang} to in effect, approximate the KS Hamiltonian at high energies by the HEG one
\begin{eqnarray}
  \hat{H}^{A} &=& -\frac{1}{2} {\bf \nabla}^2 + U_0^{HEG},
  \label{eq:H_g}
\end{eqnarray}
where $U_0^{HEG}$ is a constant potential aligning $\hat{H}^{A}$ with the system's Hamiltonian $\hat{H}^{KS}$, \eq{eq:Hamil}. In practice, in this scheme a splitting energy is chosen below which the equilibrium DM is spanned by the KS eigenfunctions and above which it is replaced by
\begin{equation}
    \label{eq:gas}
  \hat{\rho}^{A} = \left(\hat{I} + \exp\left(\frac{\hat{H}^{A} - \mu \hat{I}}{\tau_e}\right)\right)^{-1}.
\end{equation}
Since the computational cost of evaluating $\hat{\rho}^{A}$ is negligible, dramatic savings in computational cost can be achieved if the splitting energy can be pushed down to small values.

Two features of the technique described above need elaboration: (i) how best to join the two DMs $\hat{\rho}^{KS}$ and $\hat{\rho}^{A}$, and (ii), the best choice for the alignment energy $U^{HEG}_0$. A number of proposals for solving these problems have been presented in previous publications \cite{Zhang, Abinit}. However, no rigorous framework for an optimal technique has been proposed.

In the following, we derive such a framework. In order to do so, one needs to step back from \eq{eq:H_g} and instead build the spDFT formalism from bottom up, starting from a general ansatz for a DM partitioned into two spectral domains: (i) a low-energy subspace spanned by KS eigenstates $\{\psi_{{\bf k}n}\}$, and (ii) a high-energy subspace spanned by another complete set of orthonormal Bloch states $\{\psi^h_{{\bf k}n}\}$ that need not be eigenstates of the KS Hamiltonian. We then derive the expression for the free-energy functional whose variational minimium is the optimal DM that is partitioned according to the ansatz given above.  Subsequently, in \sect{sec:application} we revisit the problem of electronic structure calculations at extreme temperatures and spectral partitioning with the HEG at high energies. We derive detailed expressions for the spectral-partitioned KS Hamiltonian, forces, and stresses within both PAW and NCPP formalisms.

\subsection{Smooth spectral partitioning of the density matrix}
\label{sec:seamless}

\begin{figure}
  \centering
  \includegraphics[clip,trim=0.04cm 0.04cm 0.04cm 0.04cm, width=0.95\columnwidth]{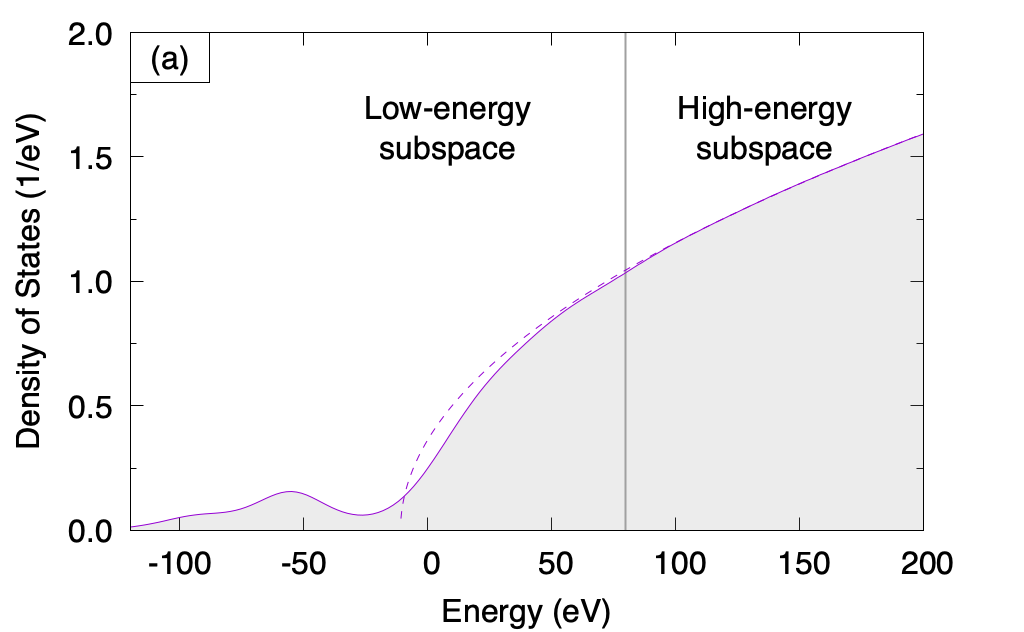}\\
    \includegraphics[clip,trim=0.04cm 0.04cm 0.04cm 0.04cm, width=0.95\columnwidth]{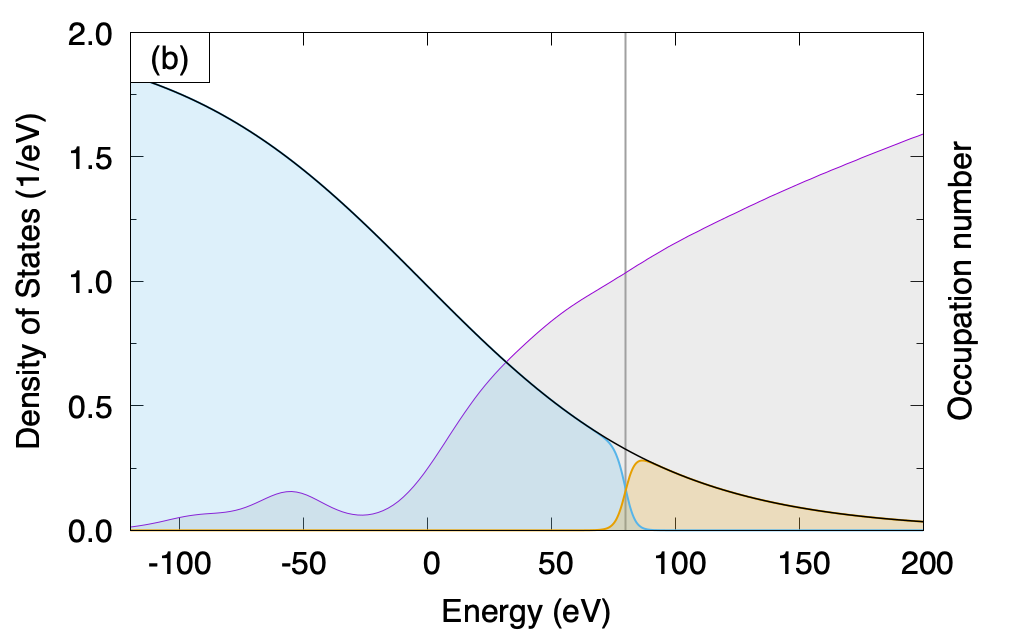}
  \caption{(a) The solid curve represents a typical electron density-of-states (DOS) of a solid and the dashed curve is the HEG DOS approximating the high-energy spectral region. (b) The  energy spectrum is decomposed into a blue region delineated by $\eta(\epsilon)$, and a yellow region bounded by $\overline{\eta}(\epsilon)$. The solid black curve depicts the electron occupation probabilities, which follow the FD distribution.}
  \label{fig:aberg}
\end{figure}

Before embarking on the derivation of the spDFT functional, we first formulate a template for the spectral partitioned equilibrium DM $\hat{\rho}^{SP}$ that results from variational minimization of this functional. In other words, our aim in this section is to specify our choice of method for joining different DM representations. For this purpose, let us consider the example in the previous subsection, where the KS Hamiltonian at high energies is approximated by $\hat{H}^A$ in \eq{eq:H_g}. Since the Bloch-wave components of the DM are mutually orthogonal, see \eq{eq:orthodm}, they can be spectrally partitioned separately and the splitting energies $\chi_{\bf k}$ are allowed to vary within the BZ. It is however, desirable that the k-dependence of the splitting energies conserve the point group symmetry of the ionic lattice and possibly the time-reversal symmetry to preserve the irreducible wedge in the BZ. 

For the following discussion, it suffices to just focus on a single Bloch-wave component $\hat{\rho}^{SP}_{\bf k}$. Preferably, the partition should consist of a smooth interpolation between $\hat{\rho}^{KS}_{\bf k}$ (\eq{eq:exact}) below a splitting energy $\chi_{\bf k}$ and an approximate $\hat{\rho}_{\bf k}^{A}$ (\eq{eq:gas}) above $\chi_{\bf k}$. For this purpose, we employ an analytic partition of unity by a function $\eta(x)$ and its complement $\overline{\eta}(x)=1-\eta(x)$, with $\eta(x)$ being a broadened step function, with $\eta(x)=1$ for $x\rightarrow-\infty$ and $\eta(x) = 0$ for $x\rightarrow\infty$. A natural choice is the sigmoid function. As a result, the SP-DM $\hat{\rho}^{SP}_{\bf k}$ becomes 
\begin{equation}
  \label{eq:rhosref}
\hat{\rho}^{SP}_{\bf k} = \eta_{\bf k}\left(\hat{H}^{KS}_{\bf k}-\mu\hat{I}\right)\hat{\rho}^{KS}_{\bf k} + \overline{\eta}_{\bf k}\left(\hat{H}^{A}_{\bf k}-\mu\hat{I}\right)\hat{\rho}^{A}_{\bf k}.
\end{equation}
with $\hat{\rho}^{KS}$ and $\hat{\rho}^{A}$ defined in Eqs.~\ref{eq:exact} and \ref{eq:gas} respectively, and
\begin{eqnarray}
  \label{eq:Q}
  \eta_{\bf k}(\hat{X_{\bf k}}) &=& \left(\hat{I} + \exp\left(\frac{\hat{X_{\bf k}} -\chi_{\bf k} \hat{I}}{\tau_s}\right)\right)^{-1}.
\end{eqnarray}
Above, $\eta_{\bf k}(\hat{H}_{\bf k}-\mu\hat{I})$ is the sigmoid function of the Hamiltonian operator relative the chemical potential, centered at the splitting energy $\chi_{\bf k}$, which can be freely chosen for each wave vector ${\bf k}$ separately. The broadening parameter $\tau_s$ can also be chosen separately for each wave vector {\bf k} but it is only a regularization parameter for enhancing numerical stability and therefore we choose to work with a single $\tau_s$ value with $\tau_s \ll \tau_e$. Figure~\ref{fig:aberg} illustrates a typical spectral splitting of the FD distribution as described above.

In the following, we address the central problem of this paper, which is to construct a variational free-energy functional whose equilibrium DM is the SP-DM $\hat{\rho}^{SP}$ in \eq{eq:rhosref}. Subsequently, we derive a general force theorem that facilitates implementation of atomic forces within various electronic structure methodologies.

\subsection{Variational spDFT free-energy functional}
\label{sec:theorem}

Consider the Hilbert space spanned by a complete set of orthonormal Bloch wavefunctions $\ket{\psi^h_{{\bf k}n}}$ subject to the same BvK boundary condition as the KS wavefunctions $\ket{\psi_{{\bf k}n}}$. Hence, as in \eq{eq:psi-u}, $\psi^h_{{\bf k}n}$ can be written in terms of functions $u^h_{{\bf k}n}$ with lattice periodicity as 
\begin{equation}
  \label{eq:phi-u}
 \psi^h_{{\bf k}n}({\bf r}) = \frac{1}{\sqrt{N_{BZ}}}  u^h_{{\bf k}n}({\bf r}) e^{i{\bf k}\cdot{\bf r}},
\end{equation}
where $u^h_{{\bf k}n}({\bf r})$ are normalized within each unit cell. The two Hilbert spaces can be transformed into one another by unitary operators $U_{{\bf k},nm}$
\begin{eqnarray}
\ket{\psi_{{\bf k}n}} &=& \sum_{m} U_{{\bf k},nm} \ket{\psi^h_{{\bf k}m}},\\
U_{{\bf k},nm} &=& \left<u^h_{{\bf k}m}|u_{{\bf k}n}\right>.
\end{eqnarray}

We define the class of Hamiltonians that are diagonal in this basis 
\begin{equation}
  \label{eq:H^h_general}
  \hat{H}^h = \sum_{{\bf k},n} \epsilon^h_{{\bf k}n} \ket{\psi^h_{{\bf k}n}}\bra{\psi^h_{{\bf k}n}}, 
\end{equation}
where $\epsilon^h_{{\bf k}n}$ are real-valued coefficients. Given any parameterization of $\hat{H}^h$, we determine the ensemble density operator $\hat{\rho}^h$ at temperature $\tau_e$ by \eq{eq:gas}, which therefore also becomes diagonal in the basis $\ket{\psi^h_{{\bf k}n}}$. However, it is important to note that contrary to \eq{eq:H_g}, we make no assumptions about the coefficients $\epsilon^h_{{\bf k}n}$, while at the same time not considering them as variational degrees of freedom. 

Our aim is to devise a variational framework whose equilibrium DM can be described by \eq{eq:rhosref}. For this purpose, we decompose each Bloch-wave component of the DM $\hat{\rho}_{\bf k}$ into two contributions $\hat{\rho}^l_{\bf k}$ and $\hat{\rho}^h_{\bf k}$, with distinct diagonal representations
\begin{eqnarray}
  \label{eq:rhos}
  \hat{\rho}^{\eta}_{\bf k} &=& \hat{\rho}^l_{\bf k} + \hat{\rho}^h_{\bf k},\\
  \label{eq:rho^l}
  \hat{\rho}^l_{\bf k} &=& \sum_{n} Q_{{\bf k}n}\ket{u_{{\bf k}n}} \bra{u_{{\bf k}n}}, \\
  \label{eq:rho^h}
\hat{\rho}^h_{\bf k} &=&\sum_{n} P_{{\bf k}n}\ket{u^h_{{\bf k}n}} \bra{u^h_{{\bf k}n}},
\end{eqnarray}
where $\{Q_{{\bf k}n}\}$, $\{P_{{\bf k}n}\}$, and $\{u_{{\bf k}n}\}$ are variational degrees of freedom. Hence, $\hat{\rho}^l_{\bf k}$ are treated fully variationally, while $\hat{\rho}^h_{\bf k}$ are only allowed to vary their orbital occupations $P_{{\bf k}n}$. Additionally, the chosen wavefunctions $u^h_{{\bf k}n}$ can in principle depend on ion positions. However, for brevity we drop this functional dependence.

In summary, our aim is to construct the spDFT free energy functional in such a way that at its variational minimum, $\rho^l$ in \eq{eq:rhos} corresponds to the first term on the right-hand side of \eq{eq:rhosref} and $\rho^h$ corresponds to the second. Hence at equilibrium, $\hat{\rho}^l$ encompasses the low-energy part of the DM and $\hat{\rho}^h$ encompasses the high-energy part.

The Helmholtz free energy of the non-interacting spectral-partitioned system in the absence of the self-consistent potential $\hat{V}_{KS}$ can now be written
\begin{equation}
  \label{eq:A^SP}
  \mathcal{A}^{SP}[\hat{\rho}^{\eta};\tau_e,\{\overline{\tau}_{\bf k}\}] = T_s[\hat{\rho}^{\eta}] - \frac{\tau_e}{N_{BZ}} \sum_{\bf k}\Tr\left\{S^{SP} \left[\hat{\rho}_{\bf k}^{\eta};\overline{\tau}_{\bf k}\right]\right\}
\end{equation}
with
\begin{equation}
  \label{eq:Dless}
  \overline{\tau}_{\bf k} = \left(\frac{\tau_s}{\tau_e},\frac{\chi_{\bf k}}{\tau_s}\right).
\end{equation}

The non-interacting kinetic energy $T_s[\hat{\rho}^{\eta}]$ in \eq{eq:A^SP} takes on the form
\begin{align}
 T_s[\hat{\rho}^{\eta}] =& -\frac{1}{2N_{BZ}}\sum_{{\bf k},n} Q_{{\bf k}n}
 \braket{u_{{\bf k}n}}{{\bf (\nabla}+i{\bf k})^2}{u_{{\bf k}n}} \nonumber \\ 
 &-\frac{1}{2N_{BZ}}\sum_{{\bf k},n} P_{{\bf k}n} \braket{u^h_{{\bf k}n}}{({\bf \nabla}+i{\bf k})^2}{ u^h_{{\bf k}n}},
\end{align}
and the spectral-partitioned entropy becomes a functional of the spectral-partitioned DM $\hat{\rho}^{\eta}$, dimensionless ratios of the splitting energies $\chi_{\bf k}$, broadening parameter $\tau_s$, and the electron temperature $\tau_e$, see \eq{eq:Dless}. Note that the SP-entropy in \eq{eq:A^SP} is decomposed into independent contributions from each Bloch-wave component of the SP-DM $\hat{\rho}^{\eta}_{\bf k}$.

\begin{widetext}
  Following the steps leading to \eq{eq:En0}, we derive the spDFT total free energy functional
  \begin{align}
    \label{eq:F_SP}
    F_{SP}[\hat{\rho};\{{\bf R}\},\tau_e,\{\overline{\tau}_{\bf k}\}] = & E_{KS}[\hat{\rho}^{\eta};\{{\bf R}\},\tau_e] - \frac{\tau_e}{N_{BZ}} \sum_{\bf k}\Tr\left\{S^{SP}\left[\hat{\rho}_{\bf k}^{\eta};\overline{\tau}_{\bf k}\right]\right\} \\
    = & \mathcal{A}^{SP}[\hat{\rho}^{\eta};\tau_e,\{\overline{\tau}_{\bf k}\}] +E_H[n^{\eta}] + F_{xc}[\hat{\rho}^{\eta},\tau_e] + \int \braket{{\bf r}}{\hat{V}_{ie}(\{{\bf R}\})}{{\bf r'}} \rho({\bf r'},{\bf r})~{\bf drdr'} \nonumber
\end{align}
where the SP charge density $n^{\eta}$ corresponds to the diagonal elements of the SP-DM $\rho^{\eta}({\bf r},{\bf r})$.
\end{widetext}

Due to the linearity of the spectral-partitioning ansatz for the DM $\hat{\rho}^{\eta}$ in \eq{eq:rhos}, functional differentiation of $E_{KS}$ in \eq{eq:F_SP} with respect to $\hat{\rho}^{\eta}$ recovers the same expression for the spectral-partitioned Hamiltonian as the unpartitioned KS Hamiltonian $\hat{H}^{KS}$ in \eq{eq:Hamil}. Therefore, at equilibrium, the spectral-partitioned energy eigenvalues can be written as
\begin{eqnarray}
  \label{eq:eexact}
  \epsilon_{{\bf k}n}^l &=& N_{BZ}\frac{\partial E_{KS}}{\partial Q_{{\bf k}n}} = \braket{u_{{\bf k}n}}{\hat{H}_{\bf k}^{KS}[\hat{\rho}^{SP}]}{u_{{\bf k}n}},\\
  \label{eq:eexact1}
  \epsilon^h_{{\bf k}n} &=& N_{BZ}\frac{\partial E_{KS}}{\partial P_{{\bf k}n}} = \braket{u^h_{{\bf k}n}}{\hat{H}_{\bf k}^{KS}[\hat{\rho}^{SP}]}{u^h_{{\bf k}n}}.
\end{eqnarray}

This is an important result. Note that the coefficients $\epsilon^h_{{\bf k}n}$ have been completely determined by the variational procedure without ever having been explicitly treated as variational degrees of freedom. In this way, the approximate Hamiltonian $\hat{H}^h$ can be determined without any prior assumptions. We see now that the spDFT framework can handle arbitrarily complex approximate Hamiltonians and that the particularly simple form of the HEG Hamiltonian in \eq{eq:H_g}, with a constant alignment potential $U_0^{HEG}$, is an exception, resulting from an intuitive ansatz, rather than the rule. In fact, we will show in section~\ref{sec:NCPP} that even for the HEG, this is too simple an assumption, and in the presence of NCPP, the HEG Hamiltonian admits a nonlocal potential.

\begin{widetext}

  We can now formulate the equilibrium spDFT free energy ${\Omega}_{SP}$, which is obtained by constrained minimization with respect to $\hat{\rho}^{\eta}$ involving the variational degrees of freedom $\{u_{{\bf k}n}\}$, $\{Q_{{\bf k}n}\}$, and $\{P_{{\bf k}n}\}$
\begin{equation}
 \label{eq:free_s}
  \Omega_{SP}[\{u^h_{{\bf k}n}\},\{{\bf R}\},\tau_e,\{\overline{\tau}_{\bf k}\}] = \min_{\{u,Q,P,\mu,\Lambda\}} ~~ F_\text{SP}[\hat{\rho}^{\eta};\{{\bf R}\},\tau_e,\{\overline{\tau}_{\bf k}\}]  
  - \mu \left(\Tr\{\hat{\rho}^{\eta}\} - N_e\right) 
  - \sum_{{\bf k},n,m} \Lambda_{nm}^{\bf k} \left(\left<u_{{\bf k}n}|u_{{\bf k}m}\right> -\delta_{nm}\right). 
\end{equation}
At equilibrium, the self-consistent SP-DM $\hat{\rho}^{SP}$ should recover \eq{eq:rhosref}.

\end{widetext}

The rest of this section will be dedicated to proving the following two theorems. 

{\bf Theorem I}. There exists an electronic entropy function $S^{SP}$ such that at the spDFT equilibrium state corresponding to the variational minimum of the spDFT total free-energy functional in \eq{eq:F_SP}, $\hat{\rho}^{SP}$ recovers the spectral partition of unity in \eq{eq:rhosref}.  

\bigskip

{\bf Theorem II}. The equilibrium spDFT free energy ${\Omega}_{SP}$ is an upper bound to the exact (unpartitioned) KS-DFT free energy, i.e. ${\Omega}_{SP}\left[\{u^h\};\{{\bf R}\},\tau_e,\{\overline{\tau}_{\bf k}\}\right] \ge {\Omega}_{KS}\left[\{{\bf R}\},\tau_e\right]$. 

\bigskip

Let us start by first proving Theorem I. For this purpose, we construct the entropy function $S^{SP}$, which at the variational minimum that defines $\Omega_{SP}$ in \eq{eq:free_s}, satisfies the following relations
\begin{eqnarray}
  \label{eq:dSdQ}
  \frac{\partial S^{SP}}{\partial Q_{{\bf k}n}} &=& \frac{\epsilon_{{\bf k}n}^l - \mu}{\tau_e},\\
  \label{eq:dSdP}
  \frac{\partial S^{SP}}{\partial P_{{\bf k}n}} &=& \frac{\epsilon^{h}_{{\bf k}n} - \mu}{\tau_e}.
\end{eqnarray}
The energy eigenvalues $\epsilon^l_{{\bf k}n}$ and $\epsilon^{h}_{{\bf k}n}$ above are defined in Eqs.~(\ref{eq:eexact}) and ~(\ref{eq:eexact1}). In order for the equilibrium SP-DM $\hat{\rho}^{SP}$ to satisfy \eq{eq:rhosref}, the solutions of Eqs.~(\ref{eq:dSdQ}) and (\ref{eq:dSdP}) should yield
\begin{eqnarray}
\label{eq:Qkn}
Q_{{\bf k}n} &=& \frac{\eta_{\bf k}\left(\epsilon^l_{{\bf k}n}-\mu\right)}{\left[1+\exp\left(\frac{\epsilon^l_{{\bf k}n} - \mu}{\tau_e}\right)\right]},\\
\label{eq:Pkn}
P_{{\bf k}n} &=& \frac{\overline{\eta}_{\bf k}\left(\epsilon_{{\bf k}n}^h-\mu\right)}{1+\exp\left(\frac{\epsilon^{h}_{{\bf k}n} - \mu}{\tau_e}\right)}
\end{eqnarray}

\begin{figure}
  \centering
  \begin{tabular}{c}
    (a)\\
    \includegraphics[clip,trim=0.04cm 0.04cm 0.04cm 0.04cm, width=0.95\columnwidth]{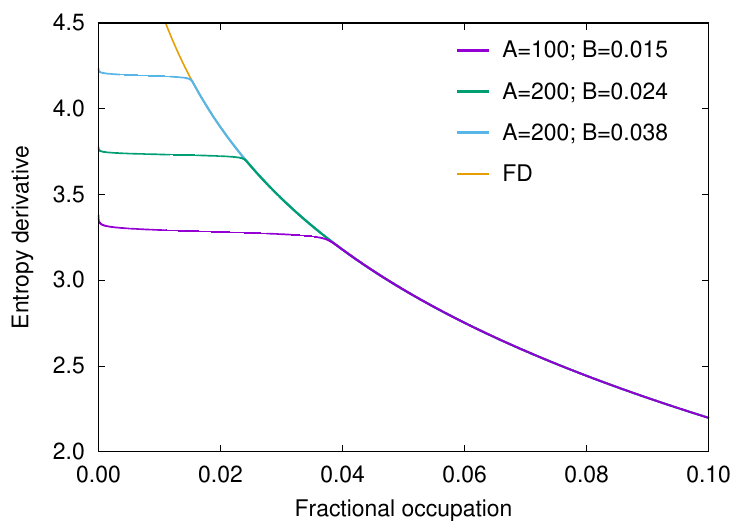}\\
    (b)\\
    \includegraphics[clip,trim=0.04cm 0.04cm 0.04cm 0.04cm, width=0.95\columnwidth]{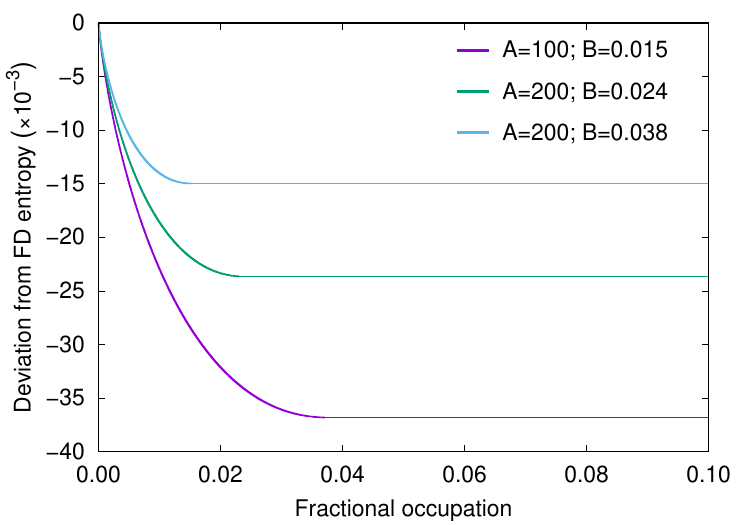}
    \end{tabular}
  \caption{(a) The entropy derivative $\dot{S}^{\eta}$,  for three different parameter sets typical of warm-dense matter applications in this paper. (b) The deviation of the SP entropy from the FD entropy, i.e. $S^{\eta}(x)-S^{FD}(x)$, as a function of occupations $x$. }
  \label{fig:entropy}
\end{figure}
    
We start by constructing an entropy function $S^{\eta}(x;\overline{\tau}_{\bf k})$ that can generate \eq{eq:Qkn} as solution. For brevity, in the following we suppress its parametric dependence and instead denote it by $S^{\eta}_{\bf k}(x)$. Hence, we make the definition 
\begin{equation}
  S^{\eta}_{\bf k}(x) \equiv S^{\eta}(x;\overline{\tau}_{\bf k}).
\end{equation}
This notation displays the ${\bf k}$-dependence of the splitting energy parameter $\chi_{\bf k}$. We thus insert \eq{eq:dSdQ} into \eq{eq:Qkn} to obtain
\begin{eqnarray}
  \label{eq:Sdot_ss}
  x &=& \frac{1}{\left[1+\exp(\dot{S}_{\bf k}^{\eta}(x))\right]\left[1+B_{\bf k}\exp\left(A\dot{S}_{\bf k}^{\eta}(x)\right)\right]},
\end{eqnarray}
with
\begin{eqnarray}
    A = \frac{\tau_e}{\tau_s}; ~~~
    B_{\bf k}= \exp\left(-\frac{\chi_{\bf k}}{\tau_s}\right); ~~~ \dot{S}_{\bf k}^{\eta} = \frac{dS_{\bf k}^{\eta}}{dx}.
\end{eqnarray}
Equation~(\ref{eq:Sdot_ss}) must be inverted to obtain $\dot{S}_{\bf k}^{\eta}(x)$. This is possible if $\dot{S}_{\bf k}^{\eta}(x)$ is a monotonic function. But \eq{eq:Sdot_ss} is a product of two sigmoid functions, each of which are separately invertible and everywhere positive. It is thus easy to see that the product must also be monotonic and thus invertible. This can be rigorously verified by examining the second derivative of the entropy with respect to occupations. For this purpose, we differentiate both sides of \eq{eq:Sdot_ss} with respect to $x$, and upon rearranging terms, a simple expression for $\ddot{S}_{\bf k}^{\eta}$ can be found:
\begin{equation}
  \label{eq:Sddot_ss}
  \ddot{S}_{\bf k}^{\eta} = -\frac{\left(1+e^{\dot{S}_{\bf k}}\right)^2\left(1+B_{\bf k}e^{A\dot{S}_{\bf k}}\right)}{e^{\dot{S}_{\bf k}} + AB_{\bf k}~e^{A\dot{S}_{\bf k}}+(A+1)B_{\bf k}e^{\dot{S}_{\bf k}+1}} < 0.
\end{equation}
From the above equation, it can be concluded that $\dot{S}_{\bf k}^{\eta}(x)$ is invertible. In the Appendix, we describe a simple procedure for calculating this function to desired accuracy. Figure~\ref{fig:entropy}(a) illustrates $\dot{S}_{\bf k}^{\eta}(x)$ for several choices of the ratios $\tau_e/\tau_s$ and $\chi_{\bf k}/\tau_s$ that are typical for applications to the warm-dense matter regime, presented later in this paper. 

It is not as straightforward to construct an entropy function that can generate \eq{eq:Pkn} as a solution. The reason is that in contrast to \eq{eq:Qkn}, \eq{eq:Pkn} is not a monotonic function of the entropy derivative defined in \eq{eq:dSdP}. This is also depicted in Fig.~\ref{fig:aberg}(b), where the $P_{{\bf k}n}$ distribution defined in \eq{eq:Pkn} is shown as the envelope of the yellow region. This distribution can instead be obtained as the difference between two monotonic functions: (i) the FD distribution, and (ii) the product of the FD distribution with the cut-off function $\eta_{\bf k}(x)$, shown as the envelope of the blue region in Fig.~\ref{fig:aberg}(b). Hence, the high-energy DM $\hat{\rho}^h$ can be considered a superposition of fictitious states with positive as well as negative occupations. As a result, the ansatz for $\hat{\rho}^{h}$ in \eq{eq:rho^h} is incomplete. Rather, it should be broken up into two contributions
\begin{eqnarray}
  \label{eq:rhos_final}
  \hat{\rho}_{\bf k}^h &=& \hat{\rho}_{\bf k}^{h+} - \hat{\rho}_{\bf k}^{h-},\\ 
  \hat{\rho}_{\bf k}^{h+} &=& \sum_{n} P^+_{{\bf k}n} \ket{u^{h}_{{\bf k}n}} \bra{u^h_{{\bf k}n}},\\
  \hat{\rho}_{\bf k}^{h-} &=& \sum_{n} P^-_{{\bf k}n} \ket{u^{h}_{{\bf k}n}} \bra{u^{h}_{{\bf k}n}},
\end{eqnarray}
where both $P^+_{{\bf k}n}$ and $P^-_{{\bf k}n}$ need now be treated as independent variational degrees of freedom. Consequently, the total spectral-partitioned entropy function $S^{SP}$ in \eq{eq:F_SP} takes the form 
\begin{equation}
  \label{eq:S^SP}
    S^{SP}[\hat{\rho}^{\eta}_{\bf k};\overline{\tau}_{\bf k}] = \sum_{\bf k} S_{\bf k}^{\eta}[\hat{\rho}^l_{\bf k}] + S^{FD}[\hat{\rho}^{h+}_{\bf k}] - S_{\bf k}^{\eta}(\hat{\rho}^{h-}_{\bf k}),
\end{equation}
where $S^{FD}(x)$ is the FD entropy defined in \eq{eq:FDentropy} and the entropy function $S_{\bf k}^{\eta}(x)$ is obtained from \eq{eq:Sdot_ss}. Following the earlier steps in this section, it is straightforward to see that upon minimization, the $Q_{{\bf k}n}$ occupations acquire the distribution in \eq{eq:Qkn} and the other ones become
\begin{eqnarray}
\label{eq:P+kn}
P^+_{{\bf k}n} &=& \frac{1}{\left[1+\exp\left(\frac{\epsilon_{{\bf k}n}^h - \mu}{\tau_e}\right)\right]},\\
\label{eq:P-kn}
P^-_{{\bf k}n} &=& \frac{\eta_{\bf k}\left(\epsilon_{{\bf k}n}^h-\mu\right)}{1+\exp\left(\frac{\epsilon^{h}_{{\bf k}n} - \mu}{\tau_e}\right)}.
\end{eqnarray}
Hence, at the variational minimum, the SP charge density can be written as
\begin{equation}
n^{SP}({\bf r}) = \sum_{{\bf k}n} Q_{{\bf k}n} |\psi_{{\bf k}n}({\bf r})|^2 + (P^+_{{\bf k}n} - P^-_{{\bf k}n}) |\psi^h_{{\bf k}n}({\bf r})|^2,
\end{equation}
with 
\begin{equation}
P^+_{{\bf k}n} - P^-_{{\bf k}n} = \frac{\overline{\eta}_{\bf k}\left(\epsilon_{{\bf k}n}^h-\mu\right)}{1+\exp\left(\frac{\epsilon^{h}_{{\bf k}n} - \mu}{\tau_e}\right)} \ge 0.
\end{equation}
It should be noted that even if $P^+_{{\bf k}n}$ and $P^-_{{\bf k}n}$ are allowed to vary independently during minimization of the right-hand side of \eq{eq:free_s}, their difference at the variational minimum is never negative, and the equilibrium SP charge density is always positive definite. The proof of Theorem I is thus complete.

We now proceed with the proof of Theorem II. Before doing so, let us define for the sake of clarity the term ``occupied domain'', which we use in the following to denote the spectral region where either one or both of $Q_{{\bf k}n}$ and $P_{{\bf k}n} = P^+_{{\bf k}n} - P^-_{{\bf k}n}$ are nonzero. 

The proof is conducted in three steps: (i) It is shown that the SP free energy $\Omega_{SP}$ in \eq{eq:free_s} becomes equal to the exact KS free energy $\Omega_{KS}$ in \eq{eq:freeen_exact}, whenever in the occupied domain, the basis functions spanning the high-energy subspace coincide with those spanning the low-energy one, $u^h_{{\bf k}n} = u_{{\bf k}n}$. (ii) Under this condition, we prove that the SP free energy is at a local minimum with respect to infinitesimal unitary transformations of $\{u^h_{{\bf k}n}\}$. (iii) We show that whenever in the occupied domain, there are basis functions such that $\left<u^h_{{\bf k}n}|u_{{\bf k}m}\right>\ne \delta_{nm}$, the SP free energy $\Omega_{SP}$ can be lowered by a unitary transformation of $\{u^h_{{\bf k}n}\}$. 

We start by proving (i). For this purpose, consider the SP free energy $\Omega_{SP}[\{u_{{\bf k}n}\}]$, where for brevity we have suppressed the dependence on the variables $\{{\bf R}\},\tau_e,\{\overline{\tau}_{\bf k}\}$. All arguments put forth below should hold for any reasonable choice of these variables. Let us now define the basis functions $w_{{\bf k}n} = u_{{\bf k}n}$ in the occupied domain. We need to show that the following relation holds
\begin{equation}
  {\Omega}_{SP}\left[\{w_{{\bf k}n}\}\right] = {\Omega}_{KS},
\end{equation}
where $\Omega_{KS}$ is the exact KS free energy in \eq{eq:freeen_exact}. This is true because $w_{{\bf k}n}$ are eigenfunctions of the KS Hamiltonian $\hat{H}_{\bf k}^{KS}$ in \eq{eq:dEdrho-k}, and therefore $\epsilon^h_{{\bf k}n} = \epsilon^l_{{\bf k}n}$ as can be concluded from Eqs.~(\ref{eq:eexact}) and (\ref{eq:eexact1}). Consequently, the occupations $P_{{\bf k}n}^- = Q_{{\bf k}n}$, see Eqs.~(\ref{eq:Qkn}) and~(\ref{eq:P-kn}), and thus the non-Fermi-Dirac contributions to the entropy function in \eq{eq:S^SP} cancel and the free energy function ${\Omega}_{KS}$ is recovered. This concludes the proof of (i). 

Next we prove (ii). For this purpose, we investigate the variations of ${\Omega}_{SP}$ with respect to the basis functions $u^h_{{\bf k}n}$ spanning the high-energy subspace. This is simplified because $\Omega_{SP}$ is already at a variational minimum with respect to the KS orbitals $u_{{\bf k}n}$ as well as the occupations $P^{\pm}_{{\bf k}n}$ and $Q_{{\bf k}n}$. Consequently, the Hellmann-Feynman theorem holds and only partial derivatives of the right-hand side of \eq{eq:free_s} with respect to $u^h_{{\bf k}n}$ contribute to the functional derivatives of ${\Omega}_{SP}$ leading to 
\begin{equation}
\label{eq:funcderiv}
\ket{g_{{\bf k}n}} = \frac{d{\Omega}_{SP}[\{w_{{\bf k}n}\}]}{d\bra{u^h_{{\bf k}n}}} = \frac{\left(P^+_{{\bf k}n}-P^-_{{\bf k}n}\right)}{N_{BZ}} \hat{H}_{\bf k}^{KS}[\hat{\rho}^{SP}]\ket{w_{{\bf k}n}}, 
\end{equation}
with the Hamiltonian $\hat{H}_{\bf k}^{KS}[\hat{\rho}^{SP}]$ defined in \eq{eq:dEdrho-k}, and the Hartree and the exchange-correlation potentials $V_H$ and $V^{xc}$ evaluated at the SP equilibrium $\hat{\rho}^{SP}$. Since we have chosen $w_{{\bf k}n} = u_{{\bf k}n}$ in the occupied domain, $\hat{\rho}^{SP}$ is equal to the exact equilibrium DM $\hat{\rho}^{KS}$, and $\hat{H}_{\bf k}^{KS}[\hat{\rho}^{KS}]$ are diagonal
\begin{equation}
\label{eq:diag}
\braket{w_{{\bf k}n}}{\hat{H}_{\bf k}^{KS}[\hat{\rho}^{KS}]}{w_{{\bf k}m}} = \delta_{nm} \epsilon^h_{{\bf k}n}.
\end{equation}

Now consider infinitesimal variations of $u^h_{{\bf k}n}$ 
\begin{eqnarray}
\ket{u^h_{{\bf k}n}} &=& \ket{u^h_{{\bf k}n}} + \alpha \ket{\delta u^h_{{\bf k}n}}, \nonumber\\
\label{eq:antiherm}
\ket{\delta u^h_{{\bf k}n}} &=& \sum_m W_{{\bf k},nm} \ket{w_{{\bf k}m}},
\end{eqnarray}

For orthonormality of the basis functions to be preserved to first order in $\alpha$, the matrices $W_{{\bf k},nm}$ must be anti-Hermitian. Hence the first-order change in energy becomes
\begin{eqnarray}
\label{eq:dOmega}
d\Omega_{SP} &=& \frac{1}{N_{BZ}}\sum_{{\bf k}n} \left<\delta u^h_{{k}n}|g_{{\bf k}n}\right> + c.c. \\ \nonumber &=& \frac{1}{N_{BZ}}\sum_{{\bf k}n} P_{{\bf k}n}\braket{\delta u^h_{{\bf k}n}}{\hat{H}_{\bf k}^{KS}[\hat{\rho}^{KS}]}{w_{{\bf k}n}} + c.c.,
\end{eqnarray}
with $P_{{\bf k}n} = P^+_{{\bf k}n}-P^-_{{\bf k}n}$. Since $\hat{H}_{\bf k}^{KS}$ is diagonal in the $w_{{\bf k}n}$ basis, see \eq{eq:diag}, the change in free energy to first order in $\alpha$ can be written as
\begin{equation}
d\Omega_{SP}^{(1)} = \frac{1}{N_{BZ}}\sum_{{\bf k}n} P_{{\bf k}n} \epsilon^h_{{\bf k}n} (W_{{\bf k},nn}+c.c.) = 0.
\end{equation}
The last equality on the right-hand side of the above equation stems from the anti-Hermitian property of the $W_{{\bf k}nm}$ matrices. This completes the proof of (ii).

Finally, we prove (iii). For this purpose, we consider $\Omega_{SP}[\{\overline{w}_{{\bf k}n}\}]$, where $\overline{w}_{{\bf k}n}$ are chosen such that within the occupied domain $\left<\overline{w}_{{\bf k}n}|u_{{\bf k}m}\right>\ne \delta_{nm}$. The derivatives of $\Omega_{SP}$ with respect to $u^h_{{\bf k}n}$ can be calculated via \eq{eq:funcderiv}, where now $\hat{\rho}^{SP}\ne \hat{\rho}^{KS}$. As a result $\braket{\overline{w}_{{\bf k}n}}{\hat{H}_{\bf k}^{KS}[\hat{\rho}^{SP}]}{\overline{w}_{{\bf k}m}}$ is not diagonal. Consequently, we can choose the anti-Hermitian matrices $W_{{\bf k},nm}$ in \eq{eq:antiherm} in such a way that the free energy $ \Omega_{SP}$ can be lowered. Following \cite{Gillan89,vasp}, we choose $W_{{\bf k},nm}$ to be
\begin{equation}
\label{eq:WWW}
W_{{\bf k},nm} = \braket{\overline{w}_{{\bf k}m}}{\hat{H}_{\bf k}^{KS}[\hat{\rho}^{SP}]}{\overline{w}_{{\bf k}n}} (P_{{\bf k}n} - P_{{\bf k}m}).
\end{equation}
Note that $W_{{\bf k},nm}$ is clearly anti-Hermitian with vanishing diagonal elements. Inserting the above ansatz for $W_{{\bf k},nm}$ into \eq{eq:antiherm} and subsequently into \eq{eq:dOmega}, we calculate the change in the free energy to first order in $\alpha$ to be 
\begin{equation}
d\Omega_{SP}^{(1)} = -2~\sum_{{\bf k},n,m} |W_{{\bf k},nm}|^2 < 0.
\end{equation}
Hence the free energy $\Omega_{SP}[\{\overline{w_{{\bf k}n}}\}]$ is not at a minimum with respect to infinitesimal unitary transformations of the basis functions $\overline{w}_{{\bf k}n}$. This concludes the proof of (iii), and completes the proof of Theorem II.  

\subsection{spDFT forces}
\label{sec:forces}

Let us start by reformulating the spDFT total-energy functional as a sum of band-structure energy and double-counting correction, which we evaluate for the self-consistent SP-DM $\hat{\rho}^{SP}$
\begin{eqnarray}
  \label{eq:E_KS}
  E_{KS}\left[\hat{\rho}^{SP},{\bf R}\right] &=& E_{bs}\left[\hat{\rho}^{SP},{\bf R}\right]  + E_{dc}[\hat{\rho}^{SP}],
\end{eqnarray}
where
\begin{eqnarray}
  \label{eq:E_bs}
  E_{bs}  &=& \frac{1}{N_{BZ}}\sum_{{\bf k}n} Q_{{\bf k}n}\epsilon^l_{{k}n} +
  \left(P^+_{{\bf k}n}-P^-_{{\bf k}n}\right)\epsilon^h_{{k}n}, \\
  E_{dc} &=&  F_{xc}[\hat{\rho}^{SP}] - \Tr\left\{\hat{V}^{xc}[\hat{\rho}^{SP}]~\hat{\rho}^{SP}\right\} - E_H[n^{SP}].~~
  \label{eq:E_dc}
\end{eqnarray}

Above $\epsilon^l_{{\bf k}n}$ and $\epsilon^h_{{\bf k}n}$ are defined by Eqs.~(\ref{eq:eexact}) and~(\ref{eq:eexact1}), with $\epsilon^l_{{\bf k}n}$ being the eigenvalues of the KS Hamiltonian $\hat{H}_{\bf k}^{KS}$ defined in \eq{eq:dEdrho-k}, with the Hartree and the XC potentials $V_H$ and $V^{xc}$ defined in \eq{eq:Vxc}. For brevity, we have dropped the explicit temperature-dependence of the XC free energy and potential in the above equations. Following Goedecker and Maschke \cite{GoedeckerMaschke}, it is easy to see that the following relation holds quite generally
\begin{equation}
  \label{eq:dEdR}
  \frac{dE_{KS}}{d{\bf R}} = \Tr\left\{\frac{\partial\hat{H}^{KS}}{\partial{\bf R}} \hat{\rho}^{SP}\right\} + \Tr\left\{\hat{H}\frac{d\hat{\rho}^{SP}}{d{\bf R}}\right\}.
\end{equation}
It is worth reiterating that $\Tr\left\{\hat{H}^{KS}\hat{\rho}^{SP}\right\} = E_{bs}$, and $\partial \hat{H}^{KS}/\partial {\bf R} = d\hat{V}_{ie}/d{\bf R}$. Hence, in the most general case we have 
\begin{eqnarray}
  \label{eq:dEdR}
  \frac{dE_{KS}}{d{\bf R}} &=& \Tr\left\{\frac{\partial\hat{H}^{KS}}{\partial{\bf R}} \hat{\rho}^l\right\} + \Tr\left\{\hat{H}^{KS}\frac{d\hat{\rho}^l}{d{\bf R}}\right\} \nonumber \\
  &+& \Tr\left\{\frac{\partial\hat{H}^{KS}}{\partial{\bf R}} \hat{\rho}^h\right\} + \Tr\left\{\hat{H}^{KS}\frac{d\hat{\rho}^h}{d{\bf R}}\right\}.
\end{eqnarray}

This is the central result of this section, which is that the atomic forces ${\bf F}_{at}$ can always be written as a sum of separate spectral contributions
\begin{eqnarray}
  \label{eq:forcedecomp}
  {\bf F}_{at} &=&   {\bf F}_{at}^l + {\bf F}_{at}^h.
\end{eqnarray}

Hence within spDFT, the contribution to forces from spectral partitions spanned by variational eigenfunctions of the KS Hamiltonian, such as ${\bf F}_{at}^l$  in the above example, are unaffected by spectral partitioning. Consequently, it is only necessary to derive and implement new expressions for ${\bf F}_{at}^h$. Further simplification can be achieved by noting that atomic forces are negative derivatives of the free energy functional ${\Omega}_{SP}$ with respect to atomic positions. Since ${\Omega}_{SP}$ is at a variational minimum with respect to the occupation numbers $P^+_{{\bf k}n}$ and $P^-_{{\bf k}n}$, we can express ${\bf F}_{at}^h$ as follows
\begin{eqnarray}
  \label{eq:F^h}
        {\bf F}_{at}^h &=& -\sum_{{\bf k}n} \frac{P^+_{{\bf k}n}-P^-_{{\bf k}n}}{N_{BZ}} \braket{u^h_{{\bf k}n}}{\frac{d\hat{V}_{ie}}{d{\bf R}}}{u^h_{{\bf k}n}} \nonumber \\
        &-&\sum_{{\bf k}n} \frac{P^+_{{\bf k}n}-P^-_{{\bf k}n}}{N_{BZ}} \left[ \braket{\frac{\partial u^h_{{\bf k}n}}{\partial {\bf R}}}{\hat{H}_{\bf k}^{KS}}{u^h_{{\bf k}n}} + c.c.\right].~~~
\end{eqnarray}

\section{\lowercase{sp}DFT Implementation}
\label{sec:implement}

In this section, we discuss implementation details of the spDFT technique for optimized performance. We will consider the same context as in the last section: two separate subspaces with the low-energy one spanned by variational KS states $\{\psi_{{\bf k}n}\}$ and the high-energy one spanned by non-variational approximate eigenstates $\{\psi^h_{{\bf k}n}\}$. The computational cost stems mainly from calculation of the low-energy variational subspace.

Before discussing optimization strategies, we first summarize the necessary steps for implementation of an spDFT scheme in an existing KS-DFT code. 

\begin{enumerate}
\item{Determine the energy eigenvalues $\epsilon^l_{{\bf k}n}$ and $\epsilon^h_{{\bf k}n}$, using Eqs.~(\ref{eq:eexact}) and (\ref{eq:eexact1}). }
\item{Calculate the occupations $Q_{{\bf k}n}$, $P^+_{{\bf k}n}$, and $P^-_{{\bf k}n}$, using Eqs.~(\ref{eq:Qkn}), (\ref{eq:P+kn}), and (\ref{eq:P-kn}). }
\item{Calculate the SP charge density using Eqs.~(\ref{eq:rhos}), (\ref{eq:rho^l}), and (\ref{eq:rho^h}).}
\item{Calculate the total SP energy by inserting the energy eigenvalues and charge density into Eqs.~(\ref{eq:E_KS}), (\ref{eq:E_bs}), and (\ref{eq:E_dc}). }
\item{Calculate the SP entropy using \eq{eq:S^SP}, with $S^{FD}(x)$ defined by \eq{eq:FDentropy}) and $S^{\eta}_{\bf k}(x)$ determined numerically by the method described in the Appendix.}
  \item{The total SP free energy is then obtained by inserting the total SP energy and SP entropy into \eq{eq:F_SP}.}
\end{enumerate}

The spDFT technique relies on the smeared cut-off function $\eta_{\bf k}(x)$ that splits the two spectral subspaces at energies $\chi_{\bf k}$ with a broadening parameter $\tau_s$. For optimal performance, an algorithm must determine the minimal number of KS states that are required to contain the low-energy subspace. However, from the perspective of practical use, it is rather desired that the number of KS states containing the low-energy subspace be chosen by the user and the algorithm determines the best set of $\chi_k$. This will be discussed in section~\ref{sec:ImplementationA} below.

In section~\ref{sec:ImplementationB}, a concern regarding the consistency of calculated spDFT free energies, forces, and stresses along paths connecting different ionic configurations, e.g., via molecular-dynamics simulations, structural relaxations, or nudged elastic band calculations, is addressed. In previous literature on HEG-extended DFT calculations of hot dense plasmas \cite{Zhang, Abinit}, such concerns have not been considered since variational free energies were not available. With the spDFT framework, one can address such issues and develop rigorous solutions for them. 

\subsection{Maximizing accuracy for given number of KS states}
\label{sec:ImplementationA}

The spDFT technique as described above is parameterized by splitting energies $\chi_{\bf k}$ and broadening width $\tau_s$. It will be shown in \sect{sec:results} that so long as $\tau_s$ is not chosen too small ($\tau_s \ll 0.1$~eV) as to slow convergence to self-consistency, and not too large ($\tau_s > 1$~eV) as to lead to suboptimal occupations of the topmost KS states, the results are insensitive to the precise value. For the cases studied in this paper and reported in \sect{sec:results}, it is found that a choice of $\tau_s$ in the range 0.1-0.2~eV works well. 

More important is the choice of the splitting energies $\chi_{\bf k}$. In general, $\chi_{\bf k}$ determine the size of the variational low-energy subspace. Hence, given any set of $\chi_{\bf k}$ values, the size of the low-energy subspace is determined by the minimal number of variational KS states beyond which the calculated total free energy $\Omega_{SP}$ remains unchanged to desired accuracy. In other words, the value of the cutoff function must vanish outside the low-energy subspace. However, since the computational cost is determined by the size of the low-energy subspace, in practice it is most straightforward for the user to determine the number of variational KS states to be included in the calculations and for the algorithm to automatically determine the optimal set of $\chi_{\bf k}$, for which the occupations of the variational KS states follow the FD distribution as closely as possible, i.e. the cutoff function $\eta_{\bf k}(x)>0$ for as many KS states as possible. 

This is easily done so long as the $\chi_{\bf k}$ are allowed to adjust during the self-consistency iterations. At each iteration, the energy eigenvalues of the KS states are calculated and sorted, from which the chemical potential $\mu$, as well as the maximal KS band energy $\epsilon_{max}({\bf k})$ at each k-point in the BZ are determined. We can now ensure that the cutoff function $\eta_{\bf k}(x)$ is only nonzero within the low-energy KS subspace by setting
\begin{equation}
  \label{eq:setchi}
  \eta_{\bf k}\left(\frac{\epsilon_{max}({\bf k})-\mu-\chi_{\bf k}}{\tau_s}\right) \approx 10^{-4}.
\end{equation}
The above relation uniquely determines $\chi_{\bf k}$ and preserves the irreducible wedge in the BZ. For the case when $\eta_{\bf k}(x)$ is the sigmoid function \eq{eq:Q}, $\chi_{\bf k}$ become
\begin{equation}
  \label{eq:setchi0}
  \chi_{\bf k} \approx \epsilon_{max}({\bf k})-\mu - 9.21\tau_s
\end{equation}

\subsection{Internal consistency along ionic trajectories}
\label{sec:ImplementationB}

The spDFT framework provides a variational formulation within which relative energies of any two ionic configurations can be evaluated. However, it also places strong constraints on the choices of spectral-partitioning parameters along ionic trajectories generated by MD simulations or structural relaxations. These constraints emerge from \eq{eq:free_s}, where the total free energy $\Omega_{SP}$ is not only a function of the ion positions $\{{\bf R}\}$, but also of the electron temperature $\tau_e$ as well as the set of all spectral-partitioning parameters, in particular the splitting energies $\{\chi_{\bf k}\}$.

In the present work, we follow a convention that has been tacitly followed in literature, which we refer to in the following as the constant-$\chi$ convention. It requires that the forces (\sect{sec:forces}) and the stresses (\sect{sec:application}) be derived by differentiation of the free energy expression \eq{eq:free_s} with respect to ionic displacements and lattice strains respectively, while holding all other parameters including the splitting energies $\chi_{\bf k}$ fixed.  Consider thus a system of ions, with the nuclei residing on sites ${\bf R}^0$. Denote the system's total spDFT free energy by ${\Omega}_{SP}^0$ and the corresponding forces by ${\bf F}^0_{at}$. A small displacement of the ions ${\bf \Delta}$ to a new position vector ${\bf R}^1 = {\bf R}^0 + {\bf \Delta}$ leads to first order in ${\bf \Delta}$ to the following change in the free energy
\begin{equation}
  \label{eq:convention}
  {\Omega}_{SP}^1 - {\Omega}_{SP}^0 = -{\bf F}^0_{at}\cdot{\bf \Delta} + \mathcal{O}(|\Delta|^2).
\end{equation}
For atomic forces ${\bf F}^0_{at}$ that are derived within the constant-$\chi$ convention above, \eq{eq:convention} strictly holds only when all $\chi_{\bf k}$ stay unchanged between the two configurations ${\bf R}^0$ and ${\bf R}^1$. Hence for MD simulations or structural relaxations guided by atomic forces and stresses that adhere to the constant-$\chi$ convention, \eq{eq:setchi} should only be used to determine $\chi_{\bf k}$ for the initial configuration. Further along any trajectory, the internal consistency between forces and free energies requires invariant $\chi_{\bf k}$ within this convention. In \sect{sec:results}, this internal consistency is examined by comparing numerical free-energy differences with analytic derivatives of the free-energy.

It should be noted that the variational formalism allows for the constant-$\chi$ convention to be abandoned for better ones depending on the application. Other conventions will require the introduction of additional terms in the expressions for analytical forces and stresses. They can be derived from concurrent differentiation of \eq{eq:free_s} with respect to both ionic displacements and splitting energies. 

We conclude this section by a brief discussion of what it means to keep $\chi_{\bf k}$ constant between two separate ionic configurations. From Eqs.~(\ref{eq:rhosref}) and (\ref{eq:Q}), it can be seen that the splitting energies are not absolute energies but are rather measured relative to the chemical potential $\mu$. The latter is a variational quantity that changes during self-consistency iterations. As a result, in the constant-$\chi$ convention, the maximum KS band energy $\epsilon_{max}({\bf k})$ that is required to be included in the calculations must be calculated from \eq{eq:setchi} or \eq{eq:setchi0}, and thus be updated concurrently with $\mu$.

\section{\lowercase{sp}DFT-HEG for high-temperature applications}
\label{sec:application}

In this section, we derive expressions for the total free energy, forces, and stresses within the spDFT-HEG scheme, assuming the XC free energy is a functional of the charge density and its gradients, and the DM is spectrally partitioned so that at low energies it is constructed from variational KS eigenstates $\{\psi_{{\bf k}n}\}$, while at high energies it is constructed from planewaves 
\begin{eqnarray}
  \label{eq:G}
  \psi^h_{{\bf k}+{\bf G}}({\bf r}) &=& \frac{1}{\sqrt{N_{BZ}}} u^{h}_{{\bf G}}({\bf r}) \exp\left(i{\bf k}\cdot {\bf r}\right), \\
\label{eq:u-G}
u^{h}_{\bf G}({\bf r}) &=& \frac{1}{\sqrt{\Omega}} \exp\left(i{\bf G}\cdot{\bf r}\right).
\end{eqnarray}
Above, the wavefunctions have been factorized into two separate sets of planewaves following \eq{eq:phi-u}. As a result, the ${\bf k}$-vectors belong to the 1st BZ, and ${\bf G}\cdot{\bf T} = 2\pi N$, where ${\bf T}$ are periodic lattice translation vectors and $N$ are integers. This decomposition is necessary for the most general implementations of the spDFT technique, when one chooses to allow the splitting energies $\chi_{\bf k}$ to vary throughout the BZ, see Eqs.~(\ref{eq:P-kn}) and (\ref{eq:setchi}). In this notation, the ansatz for the DM at high energies $\hat{\rho}^h$ becomes 
\begin{equation}
  \label{eq:rhogas}
  \rho^h({\bf r},{\bf r'}) = \frac{1}{\Omega} \sum_{{\bf k},{\bf G}} (P^+_{{\bf k}+{\bf G}} - P^-_{{\bf k}+{\bf G}})\exp\left(i({\bf k}+{\bf G})\cdot({\bf r}-{\bf r'})\right)
\end{equation}
Note that the contribution of the HEG to the charge density $\hat{\rho}^h({\bf r},{\bf r})$ is constant in space. As a result, the expression for the total SP-charge density $n^{\eta}$ becomes
\begin{eqnarray}
  \label{eq:ns}
  n^{\eta}({\bf r}) = \sum_{{\bf k}n} Q_{{\bf k}n}|\psi_{{\bf k}n}({\bf r})|^2 +
  \frac{1}{\Omega}\sum_{{\bf k},{\bf G}} (P^+_{{\bf k}+{\bf G}} - P^-_{{\bf k}+{\bf G}}) ~~~~~~
\end{eqnarray}

In the next two sections, we derive the necessary expressions for the spDFT-HEG scheme to be implemented in the two main frozen-core approaches in use today: (i) PAW, and (ii) NCPP. We assume the XC functional depends on the charge density and its gradients only, and as a result the XC potential is multiplicative and local. For brevity, we also drop the explicit temperature dependence of the XC free energy functional and the XC potential, as it neither changes the substance of the following derivations nor the final expressions. In general, implementation of the spDFT total free energy functional in a KS-DFT code requires the steps enumerated in \sect{sec:implement}. In particular, in order to implement the spDFT-HEG total free energy functional in existing PAW or NCPP codes, new expressions must be derived for the following two quantities: (i) the energy eigenvalues of the high-energy subspace $\epsilon^h_{{\bf k}n}$, and (ii) the SP-charge density $n^{\eta}({\bf r})$. The functional forms of all other quantities including the energy eigenvalues of the low-energy subspace $\epsilon^l_{{\bf k}n}$ remain unchanged. 

Regarding explicit contributions to interatomic forces from spDFT-HEG, it is clear from \eq{eq:F^h} that both terms on the right-hand side vanish, and therefore no special implementation is necessary. In contrast, there are finite contributions to macroscopic stresses from spDFT-HEG. In the following two sections, detailed derivations of these contributions will be presented.

\subsection{spDFT-HEG in the PAW method}
\label{sec:PAW}
In this section, we follow the formalism and notation of Kresse and Joubert \cite{KresseJoubert} for the PAW method. In this scheme, contrary to the NCPP formalism, the all-electron total-energy functional is in principle unchanged. Instead the valence electron wavefunctions are written in a mixed basis representation 
\begin{equation}
  \label{eq:aewave}
  \ket{\psi_{{\bf k}n}}= \ket{\tilde{\psi}_{{\bf k}n}} + \sum_{iL} \left(\ket{\phi_{iL}}-\ket{\tilde{\phi}_{iL}}\right)\ibraket{\tilde{p}_{iL}}{\tilde{\psi}_{{\bf k}n}}, 
\end{equation}
where the soft pseudo-wavefunctions $\tilde{\psi}_{{\bf k}n}$ constitute the variational degrees of freedom, and the all-electron eigenstates are recovered by a partial-wave expansion within non-overlapping augmentation spheres around each atom. $\phi_{iL}$ and  $\tilde{\phi}_{iL}$ are the atomic all-electron and pseudo-partial waves respectively, with the index $i$ enumerating atomic sites and $L$ the angular momentum channels, and the projectors $\tilde{p}_{iL}$ being dual to the pseudo-partial waves
\begin{equation}
  \ibraket{\tilde{p}_{iL}}{\tilde{\phi}_{i'L'}} = \delta_{i,i'} \delta_{L,L'}.
\end{equation}

It is important to note that for the PAW method to be an exact frozen-core scheme,  the partial-wave expansions inside the atom-centered augmentation spheres must be considered complete. We will see in \sect{sec:pawresults} that this condition can become difficult to satisfy at high electron temperatures. Nevertheless, it is straightforward to make the variational spDFT ansatz for the DM as described in \eq{eq:rhos} with $\hat{\rho}^l$ constructed from the all-electron wavefunctions in \eq{eq:aewave}, and $\hat{\rho}^h$ described by \eq{eq:rhogas}. Now following \eq{eq:ns}, the PAW SP-charge density can be written as 
\begin{equation}
  \label{eq:PAWch}
  n^{\eta}({\bf r}) = \tilde{n}({\bf r}) + n^1({\bf r}) - \tilde{n}^1({\bf r}) + n^h, 
\end{equation}
with the first term on the right-hand side being the pseudo-charge density represented on the soft planewave grid 
\begin{equation}
  \tilde{n}({\bf r}) = \sum_{{\bf k}n} Q_{{\bf k}n} \big|\tilde{\psi}_{{\bf k}n}({\bf r})\big|^2, 
\end{equation}
and the next two terms being on-site charge density contributions represented on the radial grid within each atomic augmentation sphere
\begin{eqnarray}
  n^1({\bf r}) &=& \sum_{i}\sum_{LL'} \kappa^i_{LL'} \phi_{iL}({\bf r}) \phi_{iL'}^*({\bf r})\\
  \tilde{n}^1({\bf r}) &=& \sum_{i}\sum_{LL'} \kappa^i_{LL'} \tilde{\phi}_{iL}({\bf r}) \tilde{\phi}_{iL'}^*({\bf r}),
\end{eqnarray}
with the on-site occupations $\kappa^i_{LL'}$ defined as
\begin{equation}
  \kappa^i_{LL'} = \sum_{{\bf k}n} Q_{{\bf k}n} \ibraket{\tilde{\psi}_{{\bf k}n}}{\tilde{p}_{iL}} \ibraket{\tilde{p}_{iL'}}{\tilde{\psi}_{{\bf k}n}}.
\end{equation}
The last term on the right-hand side of \eq{eq:PAWch} accounts for the HEG contribution at high energies HEG to the SP-charge density:
\begin{equation}
n^h =  \frac{1}{\Omega}\sum_{{\bf k},{\bf G}} (P^+_{{\bf k}+{\bf G}}-P^-_{{\bf k}+{\bf G}}). 
\end{equation}

By the same rationale, the non-interacting kinetic energy $T_s$ can be written as the sum of four contributions
\begin{equation}
  T_s[\hat{\rho}^{\eta}] = \tilde{T}_s[\hat{\rho}^l] + T^1_s[\hat{\rho}^l] - \tilde{T}^1_s[\hat{\rho}^l] + T^h[\hat{\rho}^h],
\end{equation}
with
\begin{eqnarray}
  \tilde{T}_s[\hat{\rho}^l] &=&   \sum_{{\bf k}n} Q_{{\bf k}n} \braket{\tilde{\psi}_{{\bf k}n}}{-\frac{\nabla^2}{2}}{\tilde{\psi}_{{\bf k}n}},\\
    T_s^1(\hat{\rho}^l) &=& \sum_{i}\sum_{LL'} \kappa^i_{LL'} \braket{\phi_{iL}}{-\frac{\nabla^2}{2}}{\phi_{iL'}},\\
    \tilde{T}_s^1(\hat{\rho}^l) &=& \sum_{i}\sum_{LL'} \kappa^i_{LL'} \braket{\tilde{\phi}_{iL}}{-\frac{\nabla^2}{2}}{\tilde{\phi}_{iL'}},
\end{eqnarray}
and
\begin{equation}
  T^h[\hat{\rho}^h] = \sum_{{\bf k},{\bf G}} (P^+_{{\bf k}+{\bf G}}-P^-_{{\bf k}+{\bf G}})\frac{({\bf k}+{\bf G})^2}{2}.
\end{equation}

Contrary to the kinetic energy term, there is more than one legitimate spDFT formulation for the interaction energy terms in PAW. For example, in the simplest implementation of spDFT-HEG within PAW, the constant charge density $n^h$ is added only to the soft pseudo-charge density $\tilde{n}({\bf r})$. This approach was taken in previous ext-FPMD implementations \cite{Zhang, Abinit}. We refer to this method as the pseudo-charge spDFT-HEG (PC-spDFT-HEG). Unfortunately, it leads to a suboptimal total-energy with systematic errors in both the exchange-correlation and the Hartree energies. Below, we analyze the PAW expressions for these interaction energies and show that the most accurate spDFT-HEG approach is obtained by adding $n^h$ to both the soft pseudo-charge density $\tilde{n}({\bf r})$, and the on-site charge densities $n^1({\bf r})$ and $\tilde{n}^1({\bf r})$. We refer to this approach as the all-electron spDFT-HEG (AE-spDFT-HEG).

\subsubsection{Exchange-correlation free energy}
Following Kresse and Joubert \cite{KresseJoubert}, the PAW XC free energy within spDFT-HEG must be written as
\begin{eqnarray}
  \label{eq:ExcPAW}
&&  F_{xc}\left[\tilde{n}+\hat{n}+\tilde{n}_c+n^h\right] + \overline{F_{xc}\left[n^1+n_c+n^h\right]} \\ && -~\overline{F_{xc}\left[\tilde{n}^1+\hat{n}+\tilde{n}_c+n^h\right]}\nonumber, 
\end{eqnarray}
where $\hat{n}$ is the compensation charge that brings the multipole moments of the on-site pseudo-charge density $\tilde{n}^1_A$ to match that of the all-electron charge density $n^1_A$, and $n_c$ and $\tilde{n}_c$ are the frozen all-electron and the partial core charge densities, respectively. Neither of the quantities $\hat{n}$, $n_c$, or $\tilde{n}_c$ are affected by the spectral partitioning.  The bars extending over the second and the third terms above denote spatial integration over atomic augmentation spheres alone. It is clear that the contribution to the exchange-correlation energy in the interstitial regions between the atomic augmentation spheres is described by the first term in \eq{eq:ExcPAW}, while within the spheres, it is the second term that determines the exchange-correlation energy with the first and third canceling. Due to the non-linearity of the exchange-correlation functional, it is thus important that all three terms in \eq{eq:ExcPAW} incorporate the constant charge density $n^h$ from the high-energy spectral region.

\subsubsection{Hartree energy}

In order to derive the correct expression for the Hartree energy, we start by the total charge density $n_T$ including the ions, the core, and the valence electrons. Following Kresse and Joubert, it is decomposed into three terms
\begin{eqnarray}
  n_T = \tilde{n}_T + n_T^1 - \tilde{n}_T^1,
\end{eqnarray}
with
\begin{eqnarray}
  \tilde{n}_T &=& \tilde{n} + \hat{n} + \tilde{n}_{Zc} + n^h,\\
  n_T^1 &=& n^1 + n_{Zc} + n^h,\label{eq:n_T1}\\
  \tilde{n}_T^1 &=& \tilde{n}^1 + \hat{n}^1 + \tilde{n}_{Zc} + n^h.\label{eq:tn_T1}
\end{eqnarray}
Above $n_{Zc}$ is the combined charge density of the ions and core electrons, and $\tilde{n}_{Zc}$ is a smooth charge distribution that coincides with $n_{Zc}$ outside the atomic core radius and have the same moment as $n_{Zc}$ inside the atomic core region. With these definitions at hand, the Hartree energy can be written~\cite{KresseJoubert}
\begin{eqnarray}
  \label{eq:hartree}
  \frac{1}{2} (n_T)(n_T) &=&   \frac{1}{2} (\tilde{n}_T)(\tilde{n}_T) +
  (n_T^1 - \tilde{n}_T^1)\tilde{n}_T \\
  &+& (n_T^1 - \tilde{n}_T^1)(n_T^1 - \tilde{n}_T^1),\nonumber
\end{eqnarray}
where we have adopted the notation from \cite{KresseJoubert}
\begin{equation}
  \label{eq:Coul}
  (a)(b) = \int \frac{a({\bf r})b({\bf r'})}{\left|{\bf r}-{\bf r'}\right|}~d{\bf r}d{\bf r'}.
\end{equation}

It is important to note that the $n_T^1-\tilde{n}_T^1$ is only nonzero inside the atomic augmentation spheres and has vanishing multipole moments due to the compensation charge $\hat{n}$, and therefore the electrostatic integrals in the second and third terms on the right-hand side of \eq{eq:hartree} have no contribution from outside the atomic augmentation sphere. As a result, \eq{eq:hartree} can be approximated by
\begin{eqnarray}
  \label{eq:hartree1}
  \frac{1}{2} (n_T)(n_T) &=&   \frac{1}{2} (\tilde{n}_T)(\tilde{n}_T) +
  \overline{(n_T^1 - \tilde{n}_T^1)\tilde{n}_T^1} \\
  &+& \overline{(n_T^1 - \tilde{n}_T^1)(n_T^1 - \tilde{n}_T^1)},\nonumber
\end{eqnarray}
where the bar extending over the second and the third terms denote the electrostatic integral only extends within the atomic augmentation spheres. Note that the factor $\tilde{n}_T$ in the second term has been replaced by $\tilde{n}_T^1$. This approximation has vanishing error whenever the partial wave expansion within the atomic augmentation spheres is complete. It also requires $\tilde{n}_T^1$ to include contribution from spectral-partitioned charges $n^h$, and thus be defined as in \eq{eq:tn_T1}. Furthermore, since $\tilde{n}_T^1$ and $n_T^1$ must have same moments, the latter must also include $n^h$ as defined in \eq{eq:n_T1}.

Starting from \eq{eq:hartree1} and reordering terms following Ref.~[\onlinecite{KresseJoubert}], the electrostatic electron-electron and electron-ion interaction energy can be reformulated as follows
\begin{eqnarray}
  \label{eq:HartreePAW}
  &&  \frac{1}{2}\left(\tilde{n}+n^h+\hat{n}\right)\left(\tilde{n}_A+n^h+\hat{n}\right) +      \\
  && \frac{1}{2}\overline{\left(n^1+n^h\right)\left(n^1+n^h\right)}  + 
  \overline{\left(n_{Zc}\right)\left(n^1+n^h\right)} - \nonumber\\
  && \frac{1}{2}\overline{\left(\tilde{n}^1+\hat{n}+n^h\right)\left(\tilde{n}^1+\hat{n}+n^h\right)} + \nonumber\\
 &&  \int V_{loc}({\bf r}) \left(\tilde{n}({\bf r})+\hat{n}({\bf r})+n^h\right)~d{\bf r} - \nonumber \\
 && \int_{\omega_a} V_{loc}({\bf r}) \left(\tilde{n}^1({\bf r})+\hat{n}({\bf r})+n^h\right)~d{\bf r} \nonumber
\end{eqnarray}
The quantity $n_{Zc}$ is the total ion and core charge density including the nuclear charge, and $V_{loc}({\bf r})$ is a local pseudopotential that outside of a core radius must be equal to the electrostatic potential from the ion and core charge $n_{Zc}$. $\omega_a$ signifies that integration is confined to within atomic augmentation spheres.

\subsubsection{Hamiltonian, forces, and stresses}
\label{sec:PAWforce}

From the preceding discussion, we can conclude that the form of the PAW total-energy functional is preserved under spectral partitioning within the spDFT-HEG scheme. This implies that the expression for the energy eigenvalues of the low-energy subspace $\epsilon^l_{{\bf k}n}$ in \eq{eq:eexact} also remain unchanged. Care must be taken to incorporate the HEG density $n^h$ into the SP-charge density. 

As for the high-energy subspace, the energy eigenvalues $\epsilon_{{\bf k}n}^h$ can be obtained by functional differentiation of the kinetic energy, exchange-correlation, and Hartree energies with respect to the occupations $P^{\pm}_{{\bf k},{\bf G}}$, see \eq{eq:eexact1}. In the following, we detail the expressions for both of AE-spDFT-HEG and PC-spDFT-HEG approaches. Starting with the AE-spDFT-HEG method, we have
\begin{equation}
  \label{eq:eps^hPAW}
  \epsilon_{{\bf k}n}^h = \sum_{{\bf k}+{\bf G}} \frac{({\bf k}+{\bf G})^2}{2} + U^{AE}_0,
\end{equation}
where $U^{AE}_0$ can be derived by differenting Eqs.~(\ref{eq:ExcPAW}) and~(\ref{eq:HartreePAW}), leading to the following expression
\begin{eqnarray}
  \label{eq:U0AEPAW}
  U^{AE}_0~\Omega &=& \int V_{xc}\left[\tilde{n}+n^h+\hat{n}+\tilde{n}_c\right]~d{\bf r} \\ &+& \int_{\omega_a} V_{xc}\left[n^1+n^h+n_c\right] d{\bf r} \nonumber\\ &-&  \int_{\omega_a} V_{xc}\left[\tilde{n}+n^h+\hat{n}+\tilde{n}_c\right]~{\bf dr} \nonumber\\
  &+&  \int_{\omega_a} V_H[n^1+n^h+n_{Zc}] - V_H[\tilde{n}^1+n^h+\hat{n}]~{\bf dr}\nonumber \\
  &-& \int_{\omega_a} V_{loc}({\bf r})~{\bf dr},\nonumber
\end{eqnarray}
where $V_{xc}[n]$ is the exchange-correlation potential, $V_H[n]$ is the Hartree potential, and $\omega_a$ signifies that integration is confined to within the atomic augmentation spheres. Note that we have dropped the contribution to $U_0^{AE}$ from the fifth term in \eq{eq:HartreePAW}, because by convention \cite{Ihm_1979}, the electron energy spectrum in standard codes is shifted by the unit-cell average of $V_{loc}$. 

Within the PC-spDFT-HEG method, the expression for the alignment potential significantly simplifies 
\begin{eqnarray}
  \epsilon_{{\bf k}n}^h &=& \sum_{{\bf k}+{\bf G}} \frac{({\bf k}+{\bf G})^2}{2} + U^{PC}_0,\\
  \label{eq:U0PCPAW}
  U^{PC}_0~\Omega &=& \int V_{xc}\left[\tilde{n}+n^h+\hat{n}+\tilde{n}_c\right]~d{\bf r}.
\end{eqnarray}

We now have all the ingredients for implementing the spDFT-HEG total free energy functional \eq{eq:F_SP}, in an existing PAW code. It is interesting to insert e.g., \eq{eq:eps^hPAW}, and \eq{eq:G} into \eq{eq:H^h_general}. Summing all the terms yields
\begin{equation}
  \hat{H}^h = -\frac{1}{2} {\bf \nabla}^2 + U_0^{AE}.
\end{equation}
Hence, without any assumptions, we have rigorously reconstructed the HEG Hamiltonian for the high-energy partition with the alignment potential $U_0^{AE}$ derived variationally without ever treating it explicitly as a variational degree of freedom. 

We now proceed with derivation of forces and stresses \cite{nielsen85,nielsen85_1}. It was shown in section~\ref{sec:forces} that in general the expression for forces can be decomposed into contributions from separate spectral regions, see \eq{eq:forcedecomp}. Also, it was concluded earlier in this section, after examining \eq{eq:F^h} that there are no additional terms associated with spDFT-HEG in the expression for forces. In contrast, we show below that there are explicit contributions to stresses. Nevertheless, since the PAW total-energy expression is preserved under both PC-spDFT-HEG and AE-spDFT-HEG approaches, the standard expressions for stress within the PAW scheme remain valid.  However, additional terms must be included: (i) a contribution from the kinetic energy of the HEG subspace to stress within both PC-spDFT-HEG and AE-spDFT-HEG approaches, and (ii) contributions to AE-spDFT-HEG stress due to incorporation of $n^h$ into the integrals of the on-site charge densities $n^1$ and $\tilde{n}^1$.

In the following, we derive these excess pressure terms, which we denote by $\Delta P^{AE}_{gas}$ and $\Delta P^{PC}_{gas}$. It should be noted that due to the uniformity of the HEG, it can only contribute explicitly to hydrostatic pressure. For clarity, we split the expressions into several terms
\begin{eqnarray}
  \label{eq:pressAEPAW}
  \Delta P^{AE}_{gas} &=& \Delta P^{kin}_{gas} + \Delta P^{xc}_{gas} + \Delta P^{H}_{gas}\\
  \label{eq:pressPCPAW}
  \Delta P^{PC}_{gas} &=& \Delta P^{kin}_{gas}
\end{eqnarray}
It is now straightforward to derive the different terms from the energy expressions above:
\begin{eqnarray}
  \label{eq:presskinPAW}
  \Delta P^{kin}_{gas} &=& \sum_{{\bf k},{\bf G}} (P^+_{{\bf k}+{\bf G}}-P^-_{{\bf k}+{\bf G}}) \frac{|{\bf k}+{\bf G}|^2}{3}\\
  \label{eq:pressxcPAW}
  \Delta P^{xc}_{gas} &=& n^h \int_{\omega_a} V_{xc}\left[n^1+n^h+n_c\right] ~d{\bf r} ~  \\ 
 &-& n^h\int_{\omega_a} V_{xc}\left[\tilde{n}^1+n^h+\hat{n}+\tilde{n}_c\right]~ {\bf dr}\nonumber\\
  \label{eq:pressHPAW}
  \Delta P^{H}_{gas} &=& n^h \int_{\omega_a} V_H[n^1+n^h+n_{Zc}]~{\bf dr} \\
  &-& n^h \int_{\omega_a} \left(~V_H[\tilde{n}^1+n^h+\hat{n}]
+ V_{loc}({\bf r})~\right)~{\bf dr} \nonumber 
\end{eqnarray}

All the pieces are now in place for implementation of spDFT-HEG within a PAW code, so long as the SP-entropy function discussed in \sect{sec:derivation} is also carefully incorporated. We will discuss in \sect{sec:results} that for standard PAW potentials, the partial-wave expansion within the atomic augmentation spheres can  become insufficiently complete at high temperatures. It is important to note that AE-spDFT-HEG (but not PC-spDFT-HEG) can alleviate this problem as the partial-wave basis set within the atomic spheres only needs to be complete for electron orbitals in the low-energy spectral region. Of course, the quality of the ansatz for the DM at high energies is crucial for the overal accuracy of the spDFT technique.

\subsection{spDFT-HEG in the NCPP method}
\label{sec:NCPP}
Separable norm-conserving pseudopotentials offer a relatively simple, accurate, and efficient formalism for removing core electrons from calculations. They replace the frozen-core all-electron Hamiltonian with an effective pseudo-Hamiltonian involving only pseudized valence electrons whose interaction with the nuclei and core electrons is described via a nonlocal pseudopotential. By far, the most popular representation for the nonlocal pseudopotential is the separable form, first proposed by Kleinman and Bylander \cite{KleinmanBylander}. This form is derived below, where for simplicity of notation, we consider a periodic unit cell of volume $\Omega$ containing $N_{at}$ atoms of only one specie. Generalization to several species is straightforward. 
\begin{eqnarray}
  \label{eq:V_KB}
  V_{ie}({\bf r},{\bf r'}) &=& \sum_R V_{loc}(r_R) \delta({\bf r}-{\bf r'}) + \\
  &&\sum_{i,l,k} c_{lk} \tilde{p}_{lk}(r_R')\tilde{p}_{lk}(r_R)\sum_m Y_{lm}({\bf \hat{r}_R'}) Y_{lm}^*({\bf \hat{r}_R}),\nonumber
\end{eqnarray}
where ${\bf r_R} = {\bf r} - {\bf R}$ with ${\bf R}$ denoting nuclear positions, the projectors $\tilde{p}_{lk}$ are radial functions localized within the atomic spheres, and the coefficients $c_{lk}$ are constants. The $l$ index enumerates the angular momentum channels, and the $k$ index enumerates the number of nonlocal projectors per $l$ channel. The local potential $V_{loc}$ is also spherically symmetric and consists of two parts
\begin{equation}
  V_{loc}(r_R) = -\frac{Z_{val}}{r_R} + V_{loc}^{nc}(r_R),
\end{equation}
where $Z_{val}$ is the pseudoatom valence charge and the second term is localized within each atomic sphere. Equation~(\ref{eq:V_KB}) does have the same structure as \eq{eq:V_ie}.

Inserting \eq{eq:u-G} into \eq{eq:eexact1} with the KS Hamiltonian defined according to Eqs.~(\ref{eq:dEdrho-k}) and ~(\ref{eq:V_KB}), the following expression for the energy eigenvalues of the HEG is obtained
\begin{equation}
  \label{eq:egas}
  \epsilon^{h}_{{\bf k}+{\bf G}} = \frac{({\bf k}+{\bf G})^2}{2} + 
  \frac{1}{\Omega}\int V_{xc}({\bf r})~d{\bf r} + V_{NL}^{FT}(|{\bf k}+{\bf G}|),
\end{equation}
where
\begin{eqnarray}
  \label{eq:Vnlq}
  V_{NL}^{FT}(q) &=& \frac{N_{at}}{\Omega}\sum_{lk}\frac{2l+1}{4\pi} \left|\tilde{p}^{FT}_{lk}(q)\right|^2,
\end{eqnarray}
with
\begin{equation}
  \tilde{p}^{FT}_{lk}(q) = \int \tilde{p}_{lk}(r)j_l(qr)~4\pi r^2~dr,
\end{equation}
where $j_l(qr)$ is the lth spherical Bessel function and $V_{NL}^{FT}(q)$ depends only on the magnitude of the planewave vector ${\bf q}$. Note that in \eq{eq:egas}, we have dropped the contribution from the non-Coulombic part of the local pseudopotential $V_{loc}^{nc}$ to $\epsilon^h_{\bf q}$, since by convention \cite{Ihm_1979}, the eigenvalue spectrum in standard codes is shifted in such a way as to exclude it.

We now have all the ingredients for implementing the spDFT-HEG total free energy functional \eq{eq:F_SP}, in an existing NCPP code. Inserting Eqs.~(\ref{eq:G}) and~(\ref{eq:egas}) into \eq{eq:H^h_general}, the effective Hamiltonian in the high-energy spectral region $\hat{H}^h$ can be expressed as 
\begin{equation}
  \label{eq:H^h}
  \hat{H}^{h} = -\frac{1}{2} {\bf \nabla}^2 + U_0^{h} + U_1^{h}(|{\bf r}-{\bf r'}|),
\end{equation}
with
\begin{eqnarray}
  \label{eq:U0h}
  U_0^{h} &=& \frac{1}{\Omega}\int V_{xc}({\bf r})~d{\bf r},\\
  U_1^{h}(r) &=& \frac{1}{\Omega}\int V_{NL}^{FT}(q) j_0(qr)~4\pi q^2~dq.
  \label{eq:U1h}
\end{eqnarray}
It is noteworthy that the expression for $\hat{H}^h$ in \eq{eq:H^h}, containing the nonlocal potential $U_1^{h}$ is more general than the intuitive ansatz (see \eq{eq:H_g}) made in previous publications \cite{Zhang, Abinit}. In fact in their original paper, Zhang {\it et al.} \cite{Zhang} note that nonlocal pseudopotentials at high energies cause an energy-dependent contribution to the potential energy, which leads to a small error if neglected. While this error is indeed small, the inclusion of \eq{eq:H^h} in its entirety is necessary for a full variational treatment. Furthermore, its incorporation introduces insignificant computational overhead. Finally, it should be emphasized that the Hamiltonian for the high-energy partition \eq{eq:H^h}, has been derived rigorously without any presumptions other than the variational principle.

Let us now proceed to discuss forces and stresses. It was shown in section~\ref{sec:forces} that in general the expression for forces can be decomposed into separate spectral contributions, see \eq{eq:forcedecomp}, and after examining \eq{eq:F^h}, it is straightforward to conclude that the contribution to forces from spDFT-HEG vanishes. In contrast, there are finite contributions to stresses. These arise from the kinetic energy, as well as the nonlocal pseudooptential $V_{NL}$ terms.

Due to the uniformity of the HEG, it can only contribute to hydrostatic pressure, and thus in the following, we derive expressions for the excess pressure $\Delta P_{gas}$ originating from these additional contributions. Hence we have
\begin{eqnarray}
  \label{eq:pressNC}
  \Delta P_{gas} = \Delta P^{kin}_{gas} + \Delta P^{NL}_{gas}.
\end{eqnarray}
The first term on the right-hand side has already been defined in \eq{eq:presskinPAW}. The second term simply follows
\begin{eqnarray}
  \label{eq:presskinNC}
  \Delta P^{NL}_{gas} &=& \frac{1}{\Omega}\sum_{{\bf k},{\bf G}} (P^+_{{\bf k}+{\bf G}}-P^+_{{\bf k}+{\bf G}})~V_{NL}^{FT}(|{\bf k}+{\bf G}|), 
\end{eqnarray}
with $V_{NL}^{FT}(q)$ defined in \eq{eq:Vnlq}. It is now straightforward to implement spDFT-HEG within the NCPP framework as long as the SP-entropy function discussed in \sect{sec:derivation} is also carefully incorporated.

\section{Application to warm- and hot-dense matter}
\label{sec:results}

In this section, we discuss electronic structure calculations in the warm- and hot-dense regimes using spDFT-HEG. We will focus on how the new variational formulation allows internally consistent free energies, forces, and stresses, while at the same time enables efficient approach to self-consistency. We will compare free energies obtained from the variational formulation \eq{eq:free_s} with those calculated based on expressions in the literature using the FD entropy \cite{Zhang, Abinit}, and discuss the consequences of inconsistency of the latter with analytic forces and stresses. We will also demonstrate that with increasing electron temperature, fewer variational KS bands are necessary to reach a given accuracy, contrary to previous findings for the ext-FPMD method \cite{Zhang, Abinit, Abinit1}.  This bodes well for the usefulness of the spDFT-HEG method for applications to high-temperature plasma. We conclude with an in-depth discussion of the accuracy of the pseudopotential formalism in general and the PAW method in particular at extreme temperatures. We will demonstrate that careful implementation of the spDFT-HEG method as outlined in section~\ref{sec:PAW} can correct some of the deficiencies of the PAW technique at plasma conditions.

The calculations presented in this section have been conducted using the VASP code \cite{vasp}, with additional implementations for the spDFT-HEG method as described above. Two systems, H and Be, are studied at elevated temperatures with all electrons present, i.e., 1 per atom for H, and 4 per atom for Be. All calculations were performed in unit cells containing a single atom using an $8\times8\times8$ $k$-point mesh for BZ integrations, and a 2000~eV planewave cutoff for H and 3000~eV for Be. The PBE parametrization \cite{PBE} of the generalized-gradient approximation to the exchange-correlation potentials was used throughout. As our purpose in this work is only to demonstrate the capabilities offered by the variational spDFT technique, we focus on just two lattice structures and densities, one for each of the two elements. We study hydrogen in a simple cubic crystal structure at a low density corresponding to a specific volume of 8~\AA $^3$/atom, and beryllium in a face-centered cubic (fcc) lattice at a relatively high density corresponding to a specific volume of 5.61~\AA $^3$/atom. 

\begin{figure}
  \centering
  \begin{tabular}{c}
    (a)\\
    \includegraphics[clip,trim=0.04cm 0.04cm 0.04cm 0.04cm, width=0.95\columnwidth]{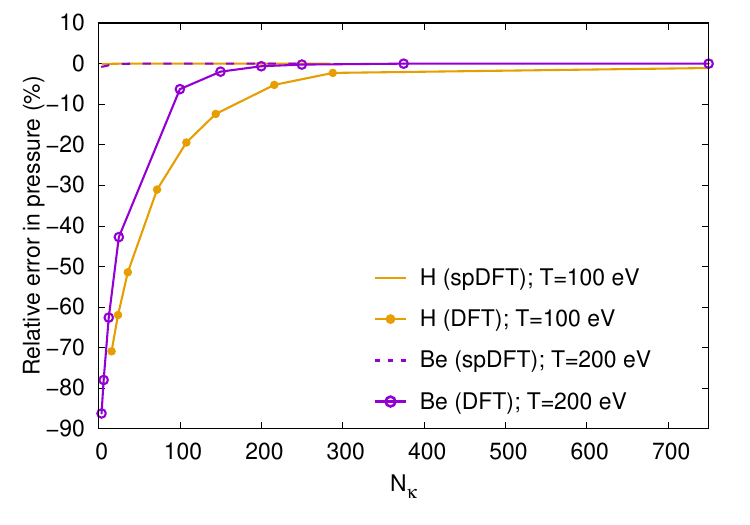}\\
    (b)\\
    \includegraphics[clip,trim=0.04cm 0.04cm 0.04cm 0.04cm, width=0.95\columnwidth]{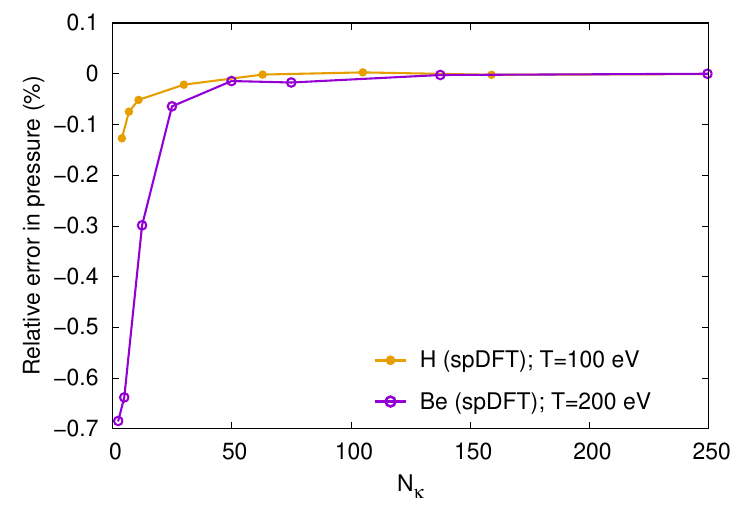}
    \end{tabular}
  \caption{Errors in calculated pressures of the H lattice at $T=100$~eV and the Be lattice at $T=200$~eV, as a function of the number of KS bands per electron $N_{\kappa}$. (a) shows comparison between spDFT-HEG and standard calculations. Note that the errors of the spDFT-HEG calculations are too small to be noticeable on the same scale with the standard technique. (b) shows comparison between the errors of the spDFT-HEG calculations of pressure for the low-density H lattice and the high-density Be lattice.}
  \label{fig:press}
\end{figure}

The most important parameter that controls the computational cost and accuracy of spDFT-HEG calculations is the number of variational KS bands that are included. Hence convergence of the calculations should be primarily investigated as a function of this number. Furthermore, it is desirable to devise a universal parameter that can be used to conduct comparative studies of the convergence of the spDFT-HEG calculations for systems with distinct chemical compositions and lattice structures. For this purpose, we introduce here $N_{\kappa}$ defined as the number of variational KS bands per electron included in an spDFT-HEG calculation. All convergence studies in the following sections will be plotted against $N_{\kappa}$.

Finally, an important technical note should be made on the particular implementations of the spDFT-HEG within PAW that have been used below. For all of the convergence studies conducted in sections {\it A} through {\it C} below, the PC-spDFT-HEG approach is utilized. While this formulation is not as accurate as AE-spDFT-HEG, it is more suitable for convergence studies as the PC-spDFT-HEG method augments only the soft pseudo-charge density $\tilde{n}({\bf r})$ with the constant charge density $n^h$ from the high-energy subspace. Since the latter can be represented with arbitrary accuracy within PAW, the convergence error of the PC-spDFT-HEG can be entirely associated with the inaccuracy of the HEG to represent the high-energy portion of the DM. Consequently, by examining the convergence of calculated pressures and free energies within PC-spDFT-HEG in sections {\it A} through {\it C} below, we can assess the efficacy of the spDFT-HEG technique in general. The AE-spDFT-HEG method, on the other hand, also augments the on-site charge densities $n^1({\bf r})$ and $\tilde{n}^1({\bf r})$ with the HEG charge density $n^h$. However, in contrast to the soft pseudo-charge density $\tilde{n}({\bf r})$, the on-site charge densities are expanded by only a few partial waves within each atomic sphere. While this expansion is nearly complete for wavefunctions in the low-energy spectral region, it becomes exceedingly inaccurate for the high-energy electron orbitals, while the HEG approximation becomes more accurate. Hence, we expect that at very high temperatures, the AE-spDFT-HEG in fact provides correction to the incompleteness of the partial-wave expansions of the on-site charge densities. We study this issue in detail in section~\ref{sec:pawresults}.

\begin{figure}
  \centering
    \includegraphics[clip,trim=0.04cm 0.04cm 0.04cm 0.04cm, width=0.95\columnwidth]{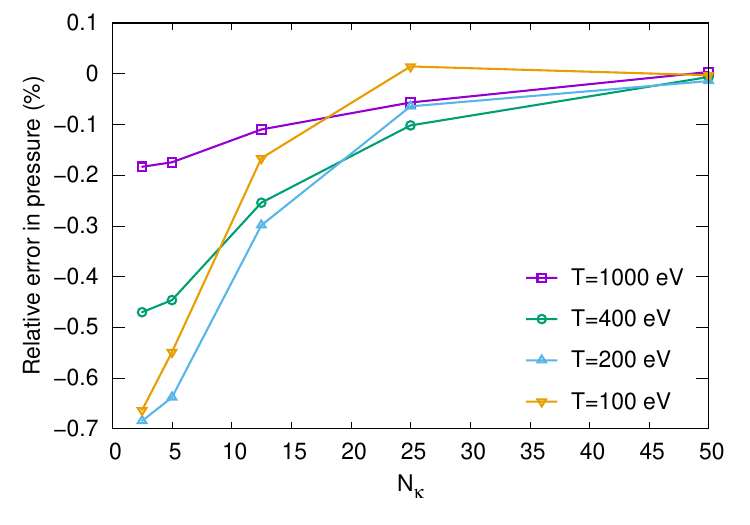} 
  \caption{Percentage error in calculated pressure at several electron temperatures, using PC-spDFT-HEG at high spectral energies, as a function of $N_{\kappa}$. The calculations use smooth spectral splitting with $\tau_s=0.2$~eV.   }
  \label{fig:pressT}
\end{figure}

\subsection{Convergence of pressure}
\label{sec:press_conv}

Figures~\ref{fig:press}(a) and (b) show the relative errors in pressure values of the H lattice at temperature 100~eV, and of the Be lattice at $T=200$~eV, calculated with and without spectral partitioning of the DM as a function of $N_{\kappa}$. The spDFT-HEG calculations are conducted using a broadening width $\tau_s = 0.2$~eV. Figure~\ref{fig:press}(a) demonstrates a dramatic improvement of the accuracy when spDFT-HEG is used to account for thermal occupations at high spectral energies. As a result, in the warm-dense regime, the computational cost of the calculations can be brought down significantly.

Note that the electron specific volume (volume per electron) of the Be lattice in this study is 1.4~\AA $^3$, which is almost six times smaller than that of the H lattice of 8~\AA $^3$. Examining Fig.~\ref{fig:press}(b), we find that the relative error in the calculated pressure of Be via the spDFT-HEG method is clearly much higher than for H. This indicates that higher densities require a larger number of KS bands per electron to reach a given accuracy. Nevertheless, the relative error of the spDFT-HEG method for Be never exceeds 1$\%$ even for $N_{\kappa}=2.5$.

The temperature dependence of the relative error in calculated pressure of the Be lattice using spDFT-HEG is shown in Fig.~\ref{fig:pressT}(a). It can be seen that between temperatures 100 eV and 1000 eV, the percentage error in pressure may be reduced by as much as 4 times. This result will be further validated in section~\ref{sec:energy_conv}, where the variational (with SP-entropy) and  non-variational (with FD-entropy) electronic free-energies as well as their convergences with $N_{\kappa}$ are compared. It will be demonstrated that, for a given $N_{\kappa}$, the free-energy error in units of thermal energy is smaller at higher temperatures, contrary to recent findings in the context of the ext-FPMD method \cite{Abinit, Abinit1, Abinit2}. This demonstrates the practical value of the rigor and consistency provided by the spDFT framework.

In conclusion, while the relative error in the calculations of pressure using spDFT-HEG diminish markedly with increasing temperature, higher densities require more KS bands to reach a given level of accuracy.

\begin{figure}
  \centering
  \begin{tabular}{c}
    (a)\\
    \includegraphics[clip,trim=0.04cm 0.04cm 0.04cm 0.04cm, width=0.95\columnwidth]{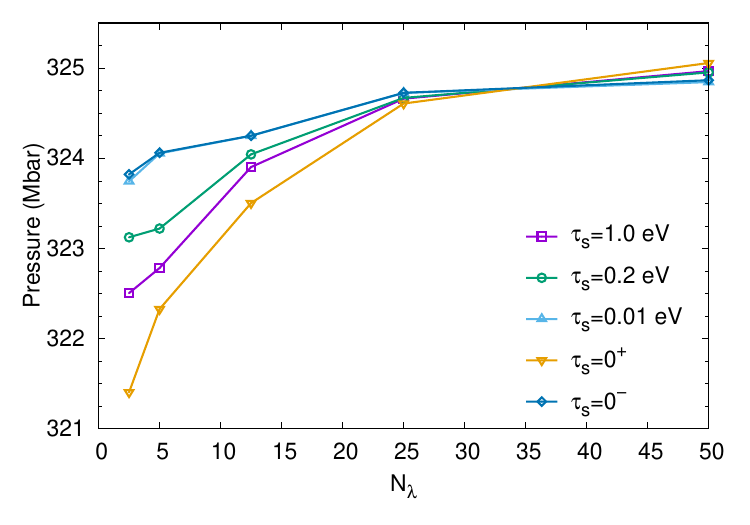}\\
    (b)\\
    \includegraphics[clip,trim=0.04cm 0.04cm 0.04cm 0.04cm, width=0.95\columnwidth]{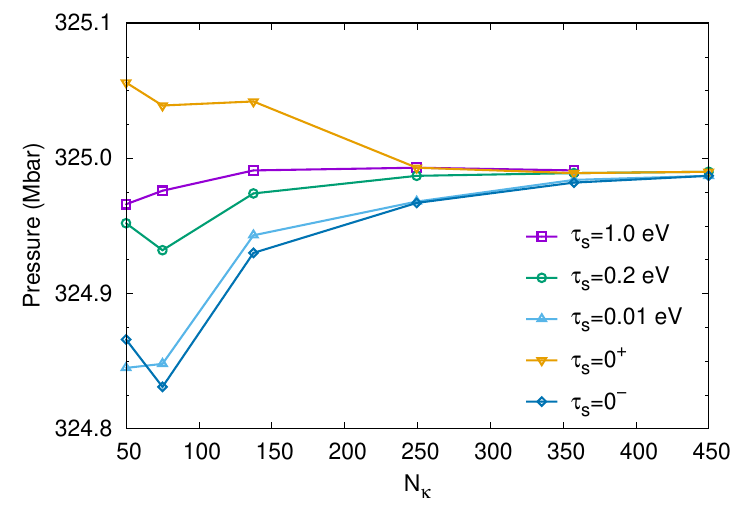}
    \end{tabular}
  \caption{Calculated pressure at $T=300$~eV as a function of $N_{\kappa}$. The different curves represent calculations with different broadening parameters $\tau_s$. (a) and (b) show convergence of pressure for two separate ranges of $N_{\kappa}$: (a) $N_{\kappa} \le 50$ and (b) $N_{\kappa} \ge 50$.}
  \label{fig:plot_tau}
\end{figure}

\subsection{spDFT-HEG with sharp versus smooth spectral splitting}
\label{sec:broadening}

In this section, we examine the relation between broadening width $\tau_s$ of the spectral splitting function, defined in \eq{eq:Q}, and convergence of pressure with respect to $N_{\kappa}$ computed within the spDFT-HEG method. We conduct a series of calculations of pressure in the Be lattice at a temperature $T=300$~eV, varying $N_{\kappa}$ from 2.5 to 450, and the broadening widths $\tau_s$ from 0.0 to 3.0~eV. The zero-broadening or sharp spectral splitting case has been included to compare the variational spDFT-HEG technique introduced in this paper, with the ext-FPMD method in the literature \cite{Zhang, Abinit}. The latter treats the shift from variational KS subspace to the HEG subspace as a sharp transition, and tacitly assumes the FD entropy applies to this situation. Certainly it does not make any sense to apply the FD entropy function to any spectral splitting of the DM other than the infinitely sharp one, i.e., when $\tau_s = 0$. The reason for this is that the FD entropy $S^{FD}$, for its definition in \eq{eq:FDentropy}, requires a diagonal representation of the DM. Hence if a DM $D$ is written as a sum of two functions $D=D_1 + D_2$, then the FD entropy associated with $D$ cannot in general be decomposed into its parts, and thus
\begin{equation}
  \label{eq:literature}
  S^{FD}(D) \ne S^{FD}(D_1) + S^{FD}(D_2),
\end{equation}
unless $D_1$ and $D_2$ operate in mutually orthogonal spectral regions. Hence, smooth transitions between subspaces as in \eq{eq:Q} require generalization beyond the FD-entropy, i.e., the SP-entropy. However, it is important to note that in nontrivial real-world applications, such as spDFT-HEG, the two subspaces on which $D_{KS}$ and $D_{HEG}$ are defined will not be strictly orthogonal and thus no matter how one splits $D$, $S^{FD}(D) \ne S^{FD}(D_{KS}) + S^{FD}(D_{HEG})$. Nevertheless, \eq{eq:literature} has been used sotto voce in the literature. As we will show, this leads to non-variational free energies that are inconsistent with analytic stresses and forces.

Figures~\ref{fig:plot_tau}(a) and (b) show the convergence of calculated pressures of the Be lattice with respect to $N_{\kappa}$, for several broadening widths $\tau_s$. It can clearly be seen that smaller $\tau_s$ are preferred if one is content with accuracies on the order of a quarter of a percent, requiring $N_{\kappa} < 20$. However, for an order of magnitude smaller errors, which require $N_{\kappa} > 50$, larger $\tau_s$ are more optimal. There are also two curves corresponding to the zero-broadening case: one marked by $\tau_s=0^-$ tracks quite closely the curve depicting pressure convergence for $\tau_s = 0.01$~eV, while the other marked by $\tau_s=0^+$ consistently exhibits larger errors than all other curves. The difference between these calculations is that in the case of $\tau_s=0^-$, the sharp spectral split occurs for each k-point of the BZ, at an energy infinitesimally smaller than the energy eigenvalue of the topmost KS band at that k-point, which as a result is left unoccupied. In contrast, in the case of $\tau_s=0^+$, the spectral splitting energy at each k-point in the BZ is infinitesimally larger than the energy eigenvalue of the topmost band at that k-point, which as a consequence is occupied according to the FD distribution. This latter case can be considered the closest to the most recent implementations of the ext-FPMD \cite{Abinit}. 

Finally, it should also be noted that we have observed slower approach to self-consistency at very small broadening widths $\tau_s \ll 0.2$~eV. In the next section, we discuss examples of situations when self-consistency can become almost impossible to achieve for $\tau_s \ll 0.2$~eV. However, we find that broadening the spectral splitting always facilitates convergence to self-consistency. For the systems and temperatures in this study, we have found that for broadening widths of $\tau_s \ge 0.1$~eV, self-consistency is reached without fail.

\begin{figure}
  \centering
  \begin{tabular}{c}
    (a)\\
    \includegraphics[clip,trim=0.04cm 0.04cm 0.04cm 0.04cm, width=0.95\columnwidth]{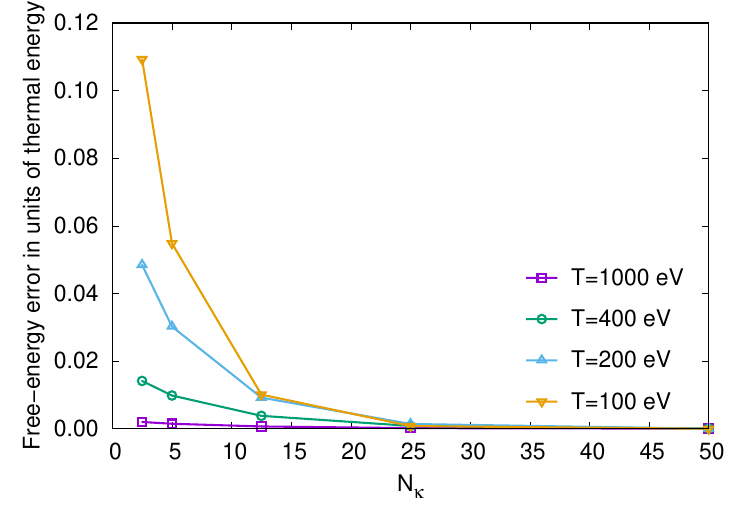}\\
    (b)\\
    \includegraphics[clip,trim=0.04cm 0.04cm 0.04cm 0.04cm, width=0.95\columnwidth]{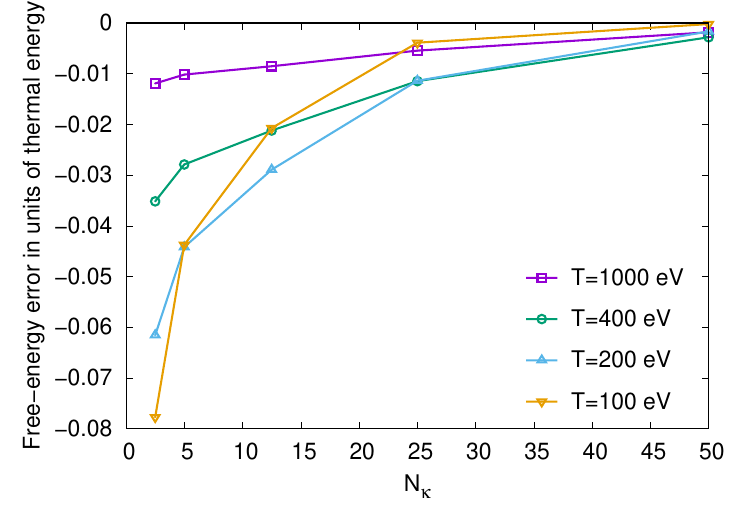}
    \end{tabular}
  \caption{Deviation in units of thermal energy per atom from the reference value of (a) the variational free energy using the SP-entropy, and (b) the non-variational free energy using the FD-entropy,  as a function of $N_{\kappa}$. }
  \label{fig:energy}
\end{figure}

\begin{figure}
  \centering
  \begin{tabular}{c}
    (a)\\
    \includegraphics[clip,trim=0.0cm 0.0cm 0.0cm 0.0cm, width=0.95\columnwidth]{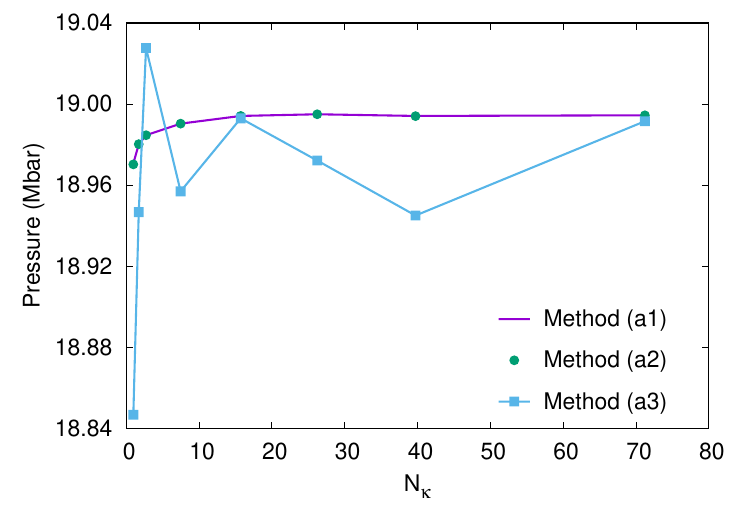}\\
    (b)\\
    \includegraphics[clip,trim=0.0cm 0.0cm 0.0cm 0.0cm, width=0.95\columnwidth]{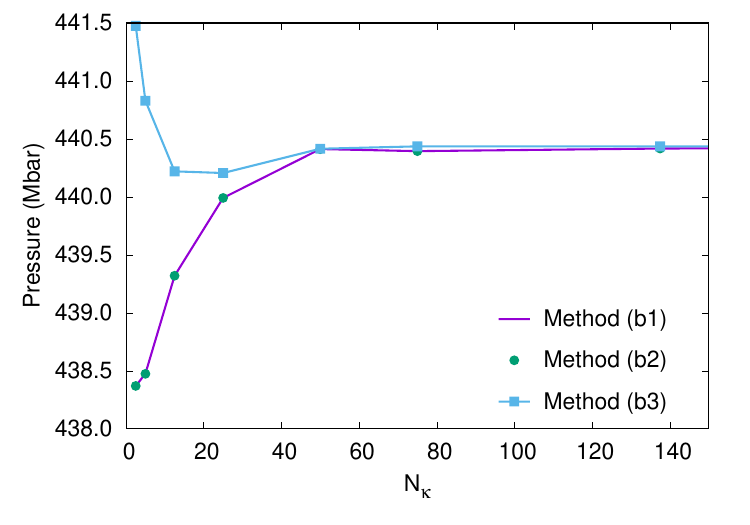}\\
    \end{tabular}
  \caption{Comparison between pressure calculated by finite differences of free energies with respect to volume change, versus analytic derivatives. (a) Comparing two finite difference schemes, one using variational and the other non-variational free energy expressions. The three methods (a1)-(a3) are desribed in the text. (b) Comparing two finite difference schemes, one using the constant-$\chi$ convention for the splitting energies at different ionic configurations, and the other using splitting energies selected from \eq{eq:setchi}, independently for each ionic configuration. The three methods (b1)-(b3) are desribed in the text. }
  \label{fig:findiff}
\end{figure}

\subsection{Convergence and consistency of the variational free energy}
\label{sec:energy_conv}

In this section we examine the variational spDFT-HEG free energy $\Omega_{SP}$ defined in \eq{eq:free_s}, and its convergence with respect to $N_{\kappa}$, as well as its consistency with the analytic pressure expression derived in \sect{sec:PAW}. We will compare with a non-variational formulation, which replaces the SP-entropy \eq{eq:S^SP} by the FD-entropy function in \eq{eq:F_SP}. Figure~\ref{fig:energy}(a) depicts the convergence with respect to $N_{\kappa}$ of $\Omega_{SP}$ in units of thermal energy per atom for the Be lattice at temperatures ranging from $100$~eV to $1000$~eV. These calculations were conducted using a broadening width $\tau_s = 0.2$~eV. Note that the variational spDFT free energy error is always positive, which is a manifestation of Theorem II proved in \sect{sec:theorem}. It states that $\Omega_{SP}$ is an upper bound to the exact free energy. In contrast, Fig.~\ref{fig:energy}(b) shows the deviation of non-variational spDFT free energies from the exact value. These calculations employ a sharp spectral splitting with $\tau_s = 0$, and use the FD-entropy function in the expression for the electronic free-energy. As a consequence, the non-variational spDFT free energies are not upper bounds to the exact value.

Closer examination of Fig.~\ref{fig:energy}(a) reveals that the convergence error in the variational free energy $\Omega_{SP}$ decreases exponentially with increasing $N_{\kappa}$. It is also clear that the magnitudes of the free energy errors relative to the thermal energies shrink with increasing temperature. For $N_{\kappa}=2.5$, the variational-spDFT free-energy error is nearly $11\%$ of the thermal energy at $T=100$~eV, but shrinks to only $0.2\%$ at $T=1000$~eV.  We can therefore conclude that, in agreement with \sect{sec:press_conv}, variational spDFT requires fewer KS eigenstates to reach a given accuracy, the higher the temperature.

We close this section by examining the consistency between variational spDFT free energy \eq{eq:free_s} and analytic pressure expressions Eqs.~(\ref{eq:pressAEPAW}) and~(\ref{eq:pressPCPAW}). This is shown in Fig.~\ref{fig:findiff}(a), where pressure versus $N_{\kappa}$ for the H lattice at $T=100$~eV is calculated in three different ways: (a1) finite differences of the variational spDFT free energies $\Omega_{SP}$ with respect to volume change, (a2) direct calculations of pressure using the analytic expression in \eq{eq:pressPCPAW}, and (a3) finite differences of non-variational free energies that use the standard FD-entropy instead of the SP-entropy. Methods (a1) and (a2) use a smooth spectral splitting with broadening width $\tau_s=0.2$~eV, while method (a3) uses a sharp spectral splitting with $\tau_s=0$. The latter approach is similar to the extended free-energy technique in \cite{Abinit}. Figure~\ref{fig:findiff}(a) clearly illustrates that methods (a1) and (a2) produce nearly indistinguishable results, while method (a3) deviates strongly from the others. To further analyze this issue, we also conducted a comparative study of the above three methods for pressures in the Be lattice at several temperatures. Again, we find the methods (a1) and (a2) nearly indistinguishable, while the finite-difference errors of method (a3) are even larger than in the case of the H lattice shown in Fig.~\ref{fig:findiff}(a). 

It is important to note that when performing finite differences of the spDFT free energies, care must be taken to adhere to the constant-$\chi$ convention, see \sect{sec:ImplementationB}. It requires keeping the splitting energies $\chi_{\bf k}$ the same for all ionic displacements. It should be borne in mind that $\chi_{\bf k}$ are measured relative to the Fermi level. The latter must be determined self-consistently for each new ionic configuration. As a result, the absolute values of the splitting energies must be adjusted concurrently. It needs to be pointed out that applying sharp spectral splitting with $\tau_s=0$ affects the approach to self-consistency. While this approach is somewhat slowed when the splitting energies are self-consistently determined according to \eq{eq:setchi}, it stalls completely when $\chi_{\bf k}$ are kept consistent with the values at another ionic configuration. In fact, the pressure calculations reported for the H lattice in Fig.~\ref{fig:findiff}(a) could not use the algorithm outlined in \sect{sec:ImplementationA} for the choice of $\chi_{\bf k}$ because the finite-difference calculations would not converge in the constant-$\chi$ mode for $\tau_s=0$. Instead, the calculations were conducted by selecting a single ${\bf k}$-independent splitting energy, which must be kept fixed throughout finite ionic displacements. Increasing the splitting energy corresponds to increasing the number of variational KS bands $N_{\kappa}$. The latter is not known beforehand. Rather it is evaluated at self-consistency by tallying the number of occupied KS bands. It should be pointed out that for sharp spectral splitting with $\tau_s=0$, even this simpler approach may fail to reach self-consistency. We thus conclude that smooth spectral splitting is necessary for robust and reliably convergent calculations. 

In order to quantify the significance of the constant-$\chi$ convention for keeping calculations along an ionic trajectory consistent, we compare in Fig.~\ref{fig:findiff}(b) three methods for computing the pressure of the Be lattice in this study at temperatures ranging from 100~eV to 1000~eV: (b1) finite differences of the variational free energy $\Omega_{SP}$ with respect to volume change, using the constant-$\chi$ convention for the splitting energies of the displaced configurations, (b2) direct calculations of pressure using the analytic expression in \eq{eq:pressPCPAW}, and (b3) finite differences of the variational free energy $\Omega_{SP}$ with respect to volume change, with the splitting energies determined independently for each volume using \eq{eq:setchi}. All calculations apply a smooth spectral splitting with the broadening width $\tau_s=0.2$~eV. Figure~\ref{fig:findiff}(b) illustrates clearly that the methods (b1) and (b2) yield indistinguishable results, while method (b3) deviates from the other two for smaller $N_{\kappa}$.

While it is important to understand the errors introduced by independent applications of \eq{eq:setchi} to different ionic configurations, one should also be cognizant that for, e.g., thermostatted molecular-dynamics simulations, the errors introduced by method (b2) may become negligible upon averaging. Also, as described in \sect{sec:ImplementationB}, constant-$\chi$ is only one of many techniques that can be used to achieve internal consistency between calculations involving different ionic configurations.  We leave further investigation of these issues for future work.

\begin{figure}
  \centering
  \begin{tabular}{c}
    (a)\\
    \includegraphics[clip,trim=0.0cm 0.0cm 0.0cm 0.0cm, width=0.95\columnwidth]{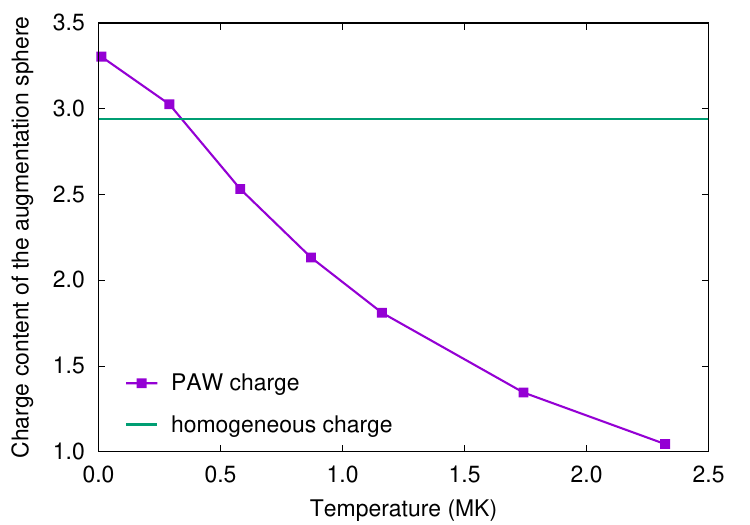}\\
    (b)\\
    \includegraphics[clip,trim=0.0cm 0.0cm 0.0cm 0.0cm, width=0.95\columnwidth]{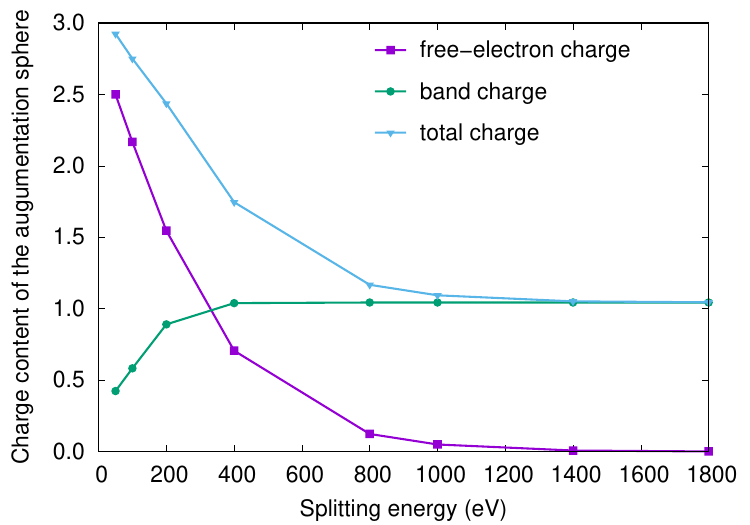}
    \end{tabular}
  \caption{(a) All-electron PAW charge in one augmentation sphere of fcc Be as a function of electron temperature. For comparison, the expected charge content of the augmentation sphere for a completely homogeneous charge distribution is also shown. (b) Breakdown of the total charge within each Be-atom augmentation sphere at 2~MK temperature into contributions from band occupations below the splitting energy and from free-electron occupations above the splitting energy. } 
  \label{fig:occ}
\end{figure}

\begin{figure}
  \centering
  \begin{tabular}{c}
    (a)\\
    \includegraphics[clip,trim=0.0cm 0.0cm 0.0cm 0.0cm, width=0.95\columnwidth]{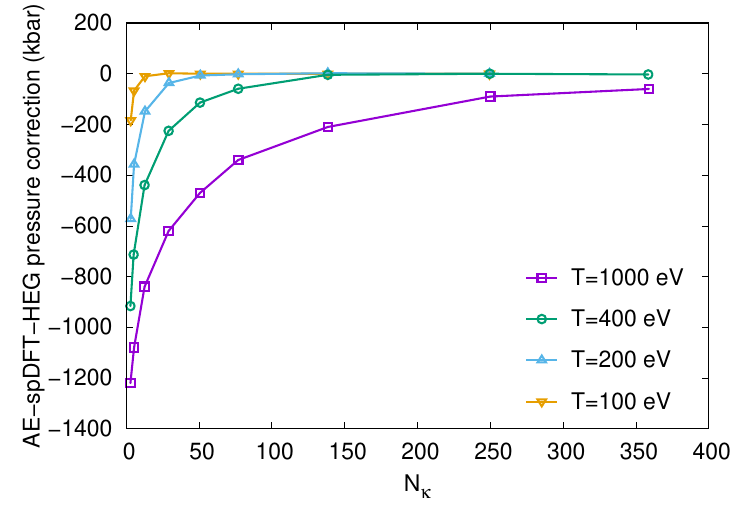}\\
    (b)\\
    \includegraphics[clip,trim=0.0cm 0.0cm 0.0cm 0.0cm, width=0.95\columnwidth]{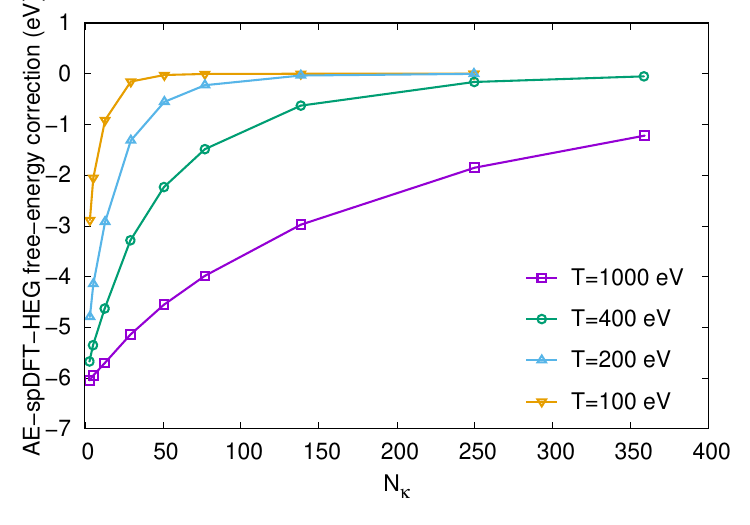}
    \end{tabular}
  \caption{AE-spDFT-HEG corrections to (a) pressures and (b) free energies calculated via PC-spDFT-HEG. }
  \label{fig:soft1cntr}
\end{figure}

\subsection{Incompleteness of the PAW basis set at high temperatures and correction by spDFT}

\label{sec:pawresults}

In \sect{sec:PAW}, we elaborated on how to best incorporate spDFT-HEG in the PAW context and developed the two approaches PC-spDFT-HEG and AE-spDFT-HEG, with the former being easier to implement but yielding less accurate PAW total free energies. Thus far, we have only presented calculations using the simpler PC-spDFT-HEG scheme, as it allows for a rigorous convergence study of the spDFT-HEG technique itself in the context of the PAW methodology. In this section, we present calculations within the AE-spDFT-HEG approach, which is always more accurate than the PC-spDFT-HEG method and no more computationally costly. The main reason we have not adopted it before is that AE-spDFT-HEG corrects some of the shortcomings of the PAW method itself, and hence it mixes errors of the spDFT-HEG method with those of the PAW. We thus now examine the corrections afforded by the AE-spDFT-HEG approach to the PC-spDFT-HEG results, which sheds light on the accuracy of standard PAW parametrizations for calculations at elevated electron temperatures.

The derivation of the PAW scheme relies on the completeness of the partial-wave expansion within the atomic augmentation spheres. However, for, e.g., PAW parametrizations used in the popular VASP program package \cite{vasp}, at most two projectors per angular-momentum channel are used for partial-wave expansion of the wavefunctions near the nuclei. It is well-known that too many nonlocal projectors can cause ghost states. This limits the ability of thus constructed PAW basis sets to represent high-energy eigenstates, which become partially occupied at high electron temperatures. This is illustrated in Fig.~\ref{fig:occ}(a), where the total charge as a function of electron temperature within one Be augmentation sphere in the fcc lattice is depicted. One normally expects an inhomogeneous charge distribution in materials with most charge concentrating near the nuclei. As temperature rises the charge distribution slowly homogenizes. However, as can be observed in Fig.~\ref{fig:occ}(a), the total charge within one Be sphere drops dramatically as temperature is increased, and at a temperature of about 2~MK, it is reduced to only 30~$\%$ of an equivalent homogeneous charge distribution. This can only be explained by the fact that overlap between the PAW nonlocal projectors and the partially occupied highly excited orbitals  become vanishingly small; or in other words, the PAW partial wave expansion within the Be atomic augmentation spheres becomes exceedingly incomplete. This has also been observed in conjunction with GW calculations within the PAW framework \cite{KresseGW}.

Figure~\ref{fig:occ}(b) illustrates the effect of spDFT-HEG on the charge content of the atomic augmentation spheres in Be. It is clearly shown that the deficiency of the PAW projectors for highly excited orbitals can be corrected in this way. However, representation of the high-energy orbitals by single planewaves does introduce other errors. Of course, as we have shown in this paper, spDFT is in no way limited to the HEG. Its strength is in its flexibility to employ the most appropriate ansatz for each spectral energy interval. The real message of Fig.~\ref{fig:occ} is that projector expansions are only valid within a finite spectral range, and outside of this range, they need to be corrected through spectral partitioning. 

Finally, we examine the corrections introduced by the AE-spDFT-HEG to the PC-spDFT-HEG of the PAW free-energy functional with the HEG at high spectral energies. Figures~\ref{fig:soft1cntr}(a) and (b) exhibit the differences in calculated pressures and free energies between the two spDFT-HEG approaches for the Be lattice at several temperatures ranging from 100~eV to 1000~eV. They show clearly that the correction magnitudes to both pressure and free energy increase when $N_{\kappa}$ is reduced. The relative improvements in pressure are quite mild, at most about 0.1$\%$ at the highest temperature $T=1000$~eV, and increase to about 0.25$\%$ as temperature is lowered to $T=100$~eV. The free-energy corrections by the AE-spDFT-HEG method are such that the AE-spDFT-HEG free energies do not monotonically increase with decreasing $N_{\kappa}$. In fact, especially at higher temperatures, the AE-spDFT-HEG free energies can be lower than fully variational PAW calculations, which is a manifestation of the incompleteness of the standard two projectors per angular-momentum channel expansions of the occupied electron orbitals within the augmentation spheres at high spectral energies.

In conclusion, use of the AE-spDFT-HEG framework in the PAW method with the HEG at high energies offers not only an accurate approximation at very low computational cost to fully variational PAW calculations of the high-temperature plasma, but also provides corrections for the incompleteness of the PAW basis set within the atomic augmentation spheres at these extreme conditions.

\section{Concluding remarks}

In this paper, we have introduced the concept of spectral partitioning of the DM in KS theory, a technique that allows for decomposition of a DM into parts, each of which specialized to describe a particular spectral domain. We have shown that given a spectral partition of unity, a variational spDFT free energy can be derived together with an entropy function associated with the chosen spectral partition. It is proven that the variational spDFT free energy is an upper bound to the exact (unpartitioned) KS-DFT free energy for the unpartitioned DM. 

The spDFT framework developed in the present work has been motivated by problems that plague calculations of equations of state of warm- and hot-dense matter. Consequently, the derivations have been within the context of finite-temperature DFT, and the Hilbert space has been decomposed into two parts: a low-energy subspace spanned by eigenfunctions of the self-consistent KS Hamiltonian, and a high-energy subspace spanned by orthogonal functions of known form, e.g., planewaves.

However, the spDFT framwork is quite general. It can be developed as well for generalized KS theories, such as hybrid functionals \cite{B3LYP,PBE0, HSE,Truhlar,Anisimov97,Coco05}. It can also be applied to ensemble-DFT functionals other than the Mermin functional, such as, e.g., one leading to Gaussian smearing of electronic occupations. Furthermore, the number of spectral intervals are not limited to two, and the variational degrees of freedom of the DM expansions in different spectral domains can be freely chosen. Hence, spectral-partitioning frameworks with arbitrary complexities can be formulated for application to matter in a wide range of conditions, from condensed matter to plasma.

\bigskip

\begin{acknowledgments}
We would like to acknowledge Philip Sterne, Sebastien Hamel and Markus D\"ane for helpful discussions. This work was performed under the auspices of the U.S. Department of Energy by Lawrence Livermore National Laboratory under Contract DE-AC52-07NA27344.
\end{acknowledgments}

\bigskip

\appendix

\section{Numerical Procedure for calculating $S^{\eta}_{\bf k}$ functions}

Equation~\ref{eq:Sdot_ss} provides a convenient numerical pathway to generate the derivative of the entropy with respect to occupation numbers $\dot{S}_{\bf k}^{\eta}$. This equation is solved independently for each ${\bf k}$-point. In this Appendix, we show a simple numerical procedure for solving it. For brevity, we drop the ${\bf k}$-indices, and focus on the following equation for $S(x)$
\begin{equation}
  \label{eq:A1}
  x = \frac{1}{\left[1+\exp(\dot{S}(x))\right]\left[1+B\exp\left(A\dot{S}(x)\right)\right]},
\end{equation}
where $A$ and $B$ are constants. Bear in mind that for the applications discussed in this paper, $A >> 1$ and $B << 1$. The range of $x$ is in the interval $[0,1]$, while the range of the derivative $\dot{S}(x)$ is unbounded, i.e. in the interval $[-\infty,\infty]$. It is easy to deduce from \eq{eq:A1}
\begin{eqnarray}
  \lim_{x\rightarrow 0}\dot{S}(x) &\rightarrow& \infty, \\
  \lim_{x\rightarrow 1}\dot{S}(x) &\rightarrow& -\infty. 
\end{eqnarray}
Taking into account that $A>>1$, then for $x\rightarrow 1$, the second factor in the denominator of \eq{eq:A1} approaches unity, and as a result, we have
\begin{equation}
  \label{eq:A2}
  \lim_{x\rightarrow 1} \dot{S}(x) = \log\left|\frac{1}{x}-1\right|,
\end{equation}
which is the same as \eq{eq:Sdot_fd} for the FD distribution. In the opposite limit, for $x\rightarrow 0$, 
\begin{equation}
  \lim_{x\rightarrow 0} \dot{S}(x) \rightarrow -\frac{\ln|Bx|}{1+A}.
\end{equation}

Integrating the above equation, one can evaluate the entropy function in the vicinity of zero occupations
\begin{equation}
  \label{eq:A3}
  \lim_{x\rightarrow 0} S(x)=\int_0^x \dot{S}(x')~dx' \rightarrow \frac{x}{1+A}\left[-\ln|Bx| + 1\right].
\end{equation}
  
All the parts are now in place for a complete algorithm for calculation of the function $S(x)$ defined in \eq{eq:A1}. We start by choosing a small number $\epsilon$, such that $\epsilon \ll 1$, which we use to determine the bounds of a closed interval for $\dot{S}$ through the conditions $ \epsilon \le B\exp\left(A\dot{S}\right)\le \frac{1}{\epsilon}$. As a result, the two bounds for this interval can be determined to be
\begin{eqnarray}
  \dot{S}_{min} &=& \frac{1}{A}\ln\left|\frac{\epsilon}{B}\right|,\\
  \dot{S}_{max} &=& -\frac{1}{A}\ln\left|\epsilon~B\right|.
\end{eqnarray}
Now, generate a uniform mesh of $\dot{S}$ values in the range $[\dot{S}_{min},\dot{S}_{max}]$ that contains $N$ elements. Next, calculate for each $\dot{S}_i$ in this set, its corresponding occupation $x_i$ using \eq{eq:A1}. Let us call $x_{1}$ the occupation corresponding to $\dot{S}_{min}$ and $x_{N}$ the occupation corresponding to $\dot{S}_{max}$. The entropy $S(x)$ at $x \le x_1$ can thus be determined from \eq{eq:A3}, while $S(x)$ in the interval $[\dot{S}_{min},\dot{S}_{max}]$ is evaluated by numerical integration. Finally the entropy at $x \ge x_N$ becomes
\begin{equation}
S(x\ge x_{N}) = S(x_{N}) - S^{FD}(x_{N}) + S^{FD}(x),
\end{equation}
where $S^{FD}(x)$ is the FD-entropy defined in \eq{eq:fFD}.

\bibliography{PR_submit}

\end{document}